\documentclass[journal=jacsat,manuscript=article]{achemso}

\usepackage{amsmath}
\usepackage{amsmath,amssymb}
\usepackage{graphicx}
\usepackage{caption}
\usepackage{color}
\usepackage{dcolumn}
\usepackage{bm}
\usepackage{float}
\usepackage{subfig}
\usepackage{comment}

\usepackage{xr}
 \author{Debasish Koner}
\affiliation{Department of Chemistry, University of Basel,
  Klingelbergstrasse 80, CH-4056 Basel, Switzerland}

\author{Max Schwilk} \affiliation{Department of Chemistry,
  University of Basel, Klingelbergstrasse 80, CH-4056 Basel,
  Switzerland}

\author{Sarbani Patra} \affiliation{Department of Chemistry,
  University of Basel, Klingelbergstrasse 80, CH-4056 Basel,
  Switzerland} 

\author{Evan J. Bieske} \affiliation{Department of Chemistry,
  University of Melbourne, Parkville 3010, Australia}

\author{Markus Meuwly} \email{m.meuwly@unibas.ch}
\affiliation[University of Basel] {Department of Chemistry, University
  of Basel,\\ Klingelbergstrasse 80, 4056 Basel, Switzerland}

\title{N$_3^+$: Full-Dimensional Potential Energy Surface, Vibrational
  Energy Levels and Ground State Dynamics}

\begin{document}

\date{\today}

\begin{abstract}
The fundamentals and higher vibrationally excited states for the
N$_3^+$ ion in its electronic ground state have been determined from
quantum bound state calculations on 3-dimensional potential energy
surfaces (PESs) at the CCSD(T)-F12 and MRCI+Q levels of theory. The
vibrational fundamentals are at 1130 cm$^{-1}$ ($\nu_1$, symmetric
stretch), 807 cm$^{-1}$ ($\nu_3$, asymmetric stretch), and 406
cm$^{-1}$ ($\nu_2$, bend) on the higher-quality CCSD(T)-F12
surface. For $\nu_1$, the calculations are close to the estimated
frequency from experiment (1170 cm$^{-1}$) and previous
calculations[Chambaud {\it et al.}, Chem. Phys. Lett., 1994, 231,
  9--12] which find it at 1190 cm$^{-1}$. Calculations of the
vibrational states on the MRCI+Q PES are in qualitative agreement with
those using the CCSD(T)-F12 PES. Analysis of the reference CASSCF wave
function for the MRCI+Q calculations provides further insight into the
shape of the PES and lends support for the reliability of Hartree-Fock
as the reference wave function for the coupled cluster
calculations. According to this, N$_3^+$ has mainly single reference
character in all low-energy regions of its electronic ground state
($^3$A$''$) 3d PES.
\end{abstract}

\section{Introduction}
N$_3^+$ is an ion involved in the chemistry of atmospheres, plasmas
and discharges. It has been found, alongside with O$_3^+$, to play an
important role in electrical discharges in air.\cite{calvaresi:2006}
Together with N$_4^+$, N$_3^+$ has been suggested to be the most
abundant nitrogen ion in the lower atmosphere of Titan where it is
formed through ternary reactions between N$^+$ and N$_2
(\Sigma_g^+$).\cite{mcewan:2000} As a consequence of its importance,
Fourier transform mass spectrometric studies of N$_3^+$ reacting with
several oxides were carried out in order to characterize the resulting
products.\cite{calvaresi:2006} The reactions are of particular
relevance when urban pollutants or greenhouse gases, such as SO$_2$,
N$_2$O, or CO$_2$ are involved, where typical reactions include
N$_3^+$+SO$_2$ $\rightarrow$ NO$^+$ + SO + N$_2$ and N$_3^+$ + N$_2$O
$\rightarrow$ NO$^+$ + 2N$_2$. In these reactions, N$_3^+$ is an N$^+$
donor participating in the general reaction scheme N$_3^+$ + XYZ
$\rightarrow$ NXYZ$^+$ + N$_2$ with subsequent decomposition of
NXYZ$^+$.  Another process relevant in Earth's atmosphere is the
charge transfer reaction N($^4$S)+N$_2^+$(X$^2 \Sigma_g^+$)
$\rightleftharpoons$ N$^+$($^3$P) + N$_2$(X$^1 \Sigma_g^+$) where
N$_3^+$ is believed to be formed through the association reaction
N$^+$ + N$_2$ $\rightarrow$ N$_3^+$.\cite{rosmus.n3:1996} Finally,
N$_3^+$ has also been implicated in the NO$^+$ production reaction
N$_3^+$ + NO $\rightarrow$ NO$^+$ + N$_2$ +
N.\cite{viggiano.n3p:2004}\\

\noindent
Experimentally, rotationally resolved A$^3 \Pi_u$ $\leftarrow$X$^3
\Sigma_g^-$ electronic spectra of N$_3^+$ have been recorded by
monitoring N$^+$ photo-products as a function of excitation wavelength
(over the 245 to 283 nm range).\cite{maier.n3p:1994} Analysis of the
ro-vibronic transitions established that in its ground electronic
state N$_3^+$ possesses a linear, centrosymmetric ($D_{\infty h}$)
structure, with an N-N bond length of 1.193 \AA\/. Although the
vibrational frequencies for the X $^3 \Sigma_g^-$ state could not be
measured directly, it was expected that they are lower than those of
the A$^3 \Pi_u$ state, which were $\nu_1 \approx 1300$ cm$^{-1}$,
$\nu_2 \approx 440$ cm$^{-1}$, and $\nu_3 \approx 1700$ cm$^{-1}$,
respectively.\cite{maier.n3p:1994} This is also consistent with a
value of $\nu_1 = 1170$ cm$^{-1}$ for the ground state of N$_3^+$,
which was deduced from photoelectron spectroscopy.\cite{dyke:1982}\\

\noindent
The first comprehensive computational study of N$_3^+$ used a complete
active space self-consistent field (CASSCF) approach and
multireference configuration interaction singles and doubles method
(MRCI) to calculate vertical excitation energies and specify the
collinear dissociation paths of the electronically excited
states.\cite{rosmus.n3:1996} Earlier attempts to theoretically
investigate the structure of the N$_{3}^{+}$ ion were inconclusive,
with studies identifying a linear N$_{3}^{+}$ molecule with either
$C_{\infty v}$ geometry\cite{archibald:1971,cai:1992} or $D_{\infty
  h}$\cite{tian:1988} symmetry for the ground state. Using highly
correlated Complete Active Space SCF followed by multireference
Averaged Coupled Pair Functional (CASSCF-ACPF) calculations it was
possible to establish the ground state $D_{\infty h}$
structure.\cite{rosmus.n3:1994} An exploratory calculation at the
MP2/aug-cc-pVTZ level of theory found that the correct symmetric
geometry with N-N bond lengths of $1.164$ \AA\/ is only obtained if
the initial structure is close to the ground state
geometry. Conversely, starting with a linear asymmetric structure, the
ground state geometry converged to one with $C_{\infty v}$ symmetry
with N-N bond lengths $r_{\rm N1-N2}=1.441$ \AA\/ and $r_{\rm
  N2-N3}=1.069$ \AA\/. This finding is also consistent with that
obtained earlier at the MRCI level of theory with a double zeta plus
polarization (DZ+P) basis set\cite{cai:1992} for which the respective
calculated bond lengths were $r_{\rm N1-N2}=1.43$ \AA\/ and $r_{\rm
  N2-N3}=1.20$ \AA\/.\\

\noindent
Vibrational calculations on a potential energy surface (PES)
calculated at the CASSCF level of theory found the vibrational
fundamentals at $\nu_1 = 1190$ cm$^{-1}$, $\nu_2 = 426$ cm$^{-1}$, and
$\nu_3 = 929$ cm$^{-1}$.\cite{rosmus.n3:1994} Subsequent {\it ab
  initio} MD simulations (10 ps) at the B3PW91/6-31G** level of theory
predicted the fundamentals at $\nu_1 = 1040$ cm$^{-1}$, $\nu_2 = 393$
cm$^{-1}$, and $\nu_3 = 900$ cm$^{-1}$ from analyzing the Fourier
transform of the velocity autocorrelation
function.\cite{jolibois:2009} These frequencies also display the
(unusual) situation that the antisymmetric stretch vibration lies
lower in frequency than the symmetric stretch. Notably, the wavenumber
of the symmetric stretch is 130 cm$^{-1}$ below the experimental
value,\cite{dyke:1982} and also considerably below that determined
from the earlier bound state calculations.\cite{rosmus.n3:1994}\\

\noindent
Here, in order to provide a more definitive characterization of the
potential energy surface for N$_3^+$ and the vibrational states
supported by it, high-level electronic structure calculations are
combined with state-of-the art methods to represent 3-dimensional PESs
and quantum bound state calculations. This work is also a prerequisite
for investigating the photodissociation dynamics for the excited
states of N$_3^+$.\\

\section{Computational Methods}
First the generation and representation of the 3-dimensional PES for
N$_3^+$ is described. This PES is then used for computing the
vibrational bound states using methods to solve the time-dependent and
time-independent nuclear Schr\"odinger equation.

\subsection{The Ground State PES for N$_{3}^{+}$}
All electronic structure calculations were performed using the Molpro
2019.1\cite{molpro,MOLPRO_brief} software. The PES for N$_{3}^{+}$ has
been determined at two different levels of theory. First, calculations
at the
CCSD(T)-F12b\cite{knizia:2009}/aug-cc-pVTZ-f12\cite{Kirk:VnZF1208}
level with the corresponding density fitting and a resolution of the
identity basis sets in the F12 computation provide a high quality
ground state PES. A spin-restricted open-shell Hartree-Fock reference
wave function was used for the spin unrestricted coupled cluster wave
function. Secondly,
MRCI\cite{wer88:5803,kno88:514}/aug-cc-pVTZ\cite{dun89:1007} with the
Davidson quadruples correction\cite{davidson:1974} (MRCI+Q)
calculations with a
CASSCF\cite{wen85:5053,kno85:259,wer80:2342,werner:2019} reference
wave function were carried out in view of future explorations of the
photo-dissociation process which require fully-dimensional PESs for
electronically excited states.\\

\noindent
For CASSCF the active space was the full valence space and the
calculations were performed as state-averaged calculations including
the two lowest singlet and triplet spin states of A$'$ and A$''$
symmetry, respectively. Multi-reference effects are potentially
important in certain regions of the PES. For this, an analysis of the
CASSCF wave function provides additional information about the
reliability of the Hartree-Fock reference wave function for the
coupled cluster calculations. Furthermore, for the dissociation
N$_3^+$ $\rightarrow$ N$_2$ + N$^+$ the role of the energetically
adjacent N$_2^+$ + N state can also be discussed based on analyzing
the CASSCF wave function. \\

\begin{figure}[htbp]
\includegraphics[width=0.3\textwidth]{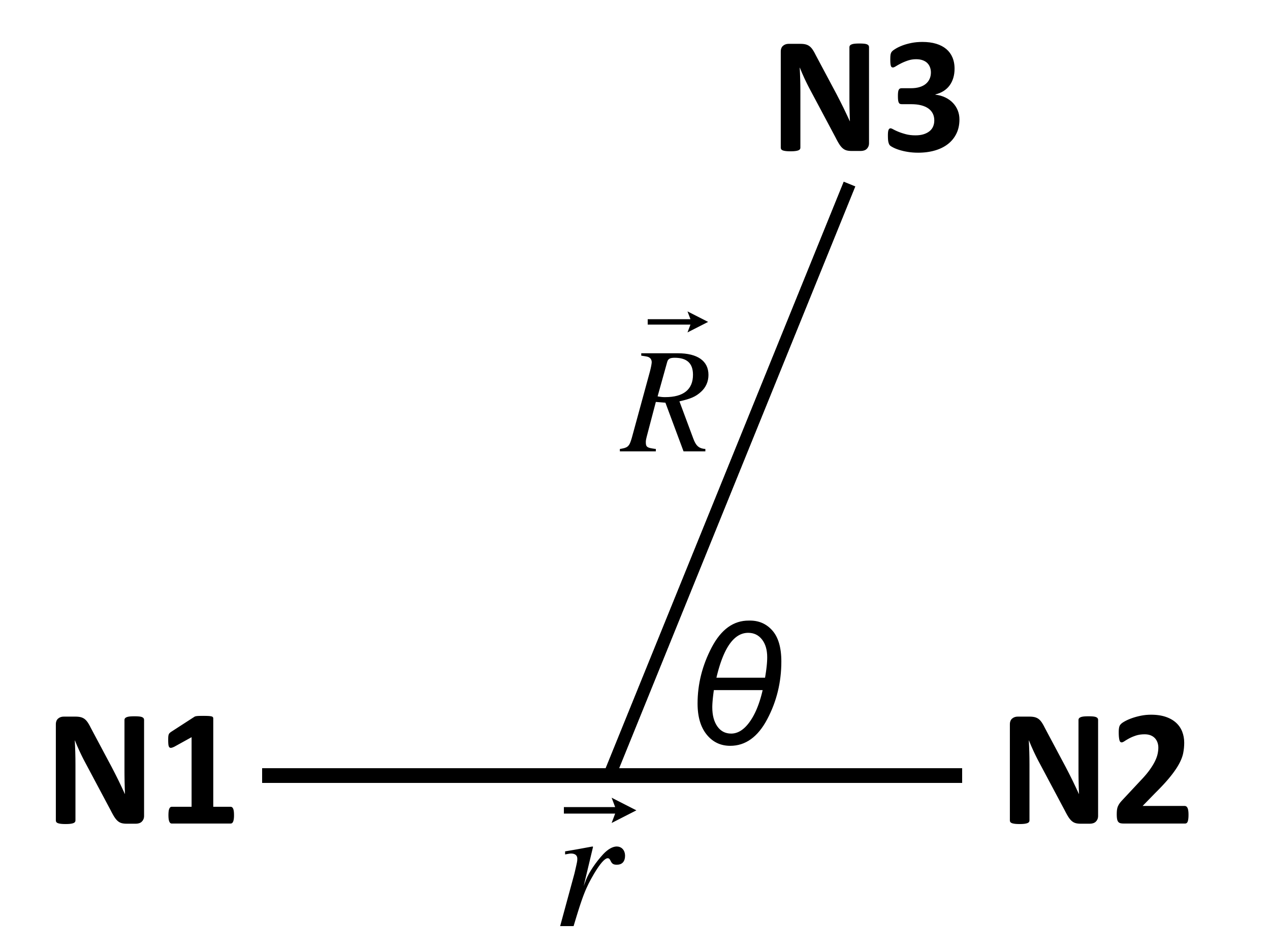}
\caption{The coordinate system $(R,r,\theta)$ used for the
  3-dimensional PES for N$_3^+$. The vector $\vec{r}$ is the N1--N2
  separation, $\vec{R}$ is the separation between N3 and the center of
  mass of N1--N2, and $\theta$ is the angle between the two distance
  vectors.}
\label{fig:n3pcoord}	
\end{figure}

\noindent
Jacobi coordinates $(R, r, \theta)$ were used to define the molecular
geometry. Here, $r$ is the separation between nitrogen N1 and N2, $R$
is the distance between N3 and the center of mass of N1 and N2, and
$\theta$ is the angle between $\vec{r}$ and $\vec{R}$, see Figure
\ref{fig:n3pcoord}. The angular grid is defined by Gauss-Legendre
quadrature points between 0 and $90^\circ$ considering the symmetry of
the system. Details of the angular and radial grids for the
CCSD(T)-F12b and MRCI+Q calculations are given in Table
\ref{sitab:triat-grid}.\\

\noindent
In addition, the PES of N$_{2}$ is required in order to correctly
describe the long-range asymptotic behaviour. Calculations of the
N$_{2}$ energies were carried out at the same level of theory as for
the underlying 3-dimensional PESs. The 1-dimensional curve for N$_2$
was then represented by a 1-dimensional reproducing kernel Hilbert
space (RKHS, see below). With this, the energy of N$_3^+$ is
\begin{equation}
V(R,r,\theta)=E(R,r,\theta)+E_{\rm N_2}(r)
\end{equation}
where $E(R,r,\theta)$ is the interaction energy of the N$^+$+N$_2$
system and $E_{\rm N_2}(r)$ is the energy of the N$_2$ molecule
(N1-N2).\\

\subsection{Reproducing Kernel Representation of the 3d-PES for N$_{3}^{+}$}
To represent the 3- and 1-dimensional energies $E(R,r,\theta)$ and
$E_{\rm N_2}(r)$ a reproducing kernel Hilbert space (RKHS)
interpolation scheme is used.\cite{rabitz:1996,MM.rkhs:2017} Starting
from $N$ known values $f({\bf x}_i)$ for coordinates ${\bf x}_i$, the
RKHS theorem states that for arbitrary values ${\bf x}$ the function
$f({\bf x})$ can be evaluated as a linear combinations of kernel
products
\begin{equation}
f({\bf{x}}) = \sum_{i=1}^{N} \alpha_i K ({\bf{x}}, {\bf{x}}_i),
\end{equation}
where $\alpha_i$ are the coefficients and $K({\bf{x}}, {\bf{x}}_i)$
are the reproducing kernels.  The coefficients can be evaluated from
the known values $f({\bf x}_i)$ by solving a set of linear equations
\begin{equation}
 f({\bf{x}}_j)= \sum_{i=1}^{N} \alpha_i K ({\bf{x}}_i, {\bf{x}}_j).
\end{equation}
Thus, a RKHS exactly reproduces the input data at the reference points
${\bf x}_i$. However, for noisy data a small damping value is added to
the diagonal elements to regularize the data set. The derivatives of
$f({\bf{x}})$ can be calculated analytically from the kernel functions
$K({\bf{x}}, {\bf{x}}')$.  For a multidimensional function the
$D$-dimensional kernel can be constructed as the product of $D$
1-dimensional kernels $k(x, x')$
\begin{equation}
 K({\bf{x}}, {\bf{x}}') = \prod_{d=1}^{D}k^{(d)}(x^{(d)}, x'^{(d)}),
\end{equation}
where $k^{(d)}(x^{(d)}, x'^{(d)})$ are the 1-dimensional kernels for
the $d$-th dimension.\\

\noindent
In the present work, the explicit form of the reciprocal power decay
kernel polynomials are used for the radial coordinates. Kernel
functions ($k^{[n,m]}$) with smoothness $n=2$ and asymptotic decay $m=4$
are used for the $R$ dimension
\begin{equation}
\label{k24}
 k^{[2,4]}(x,x') = \frac{2}{15}\frac{1}{x^5_{>}} -
 \frac{2}{21}\frac{x_<}{x^6_>},
\end{equation}
while $n=2$ and $m=6$ is used for the $r$ dimension
\begin{equation}
 k^{[2,6]}(x,x') = \frac{1}{14}\frac{1}{x^7_{>}} - \frac{1}{18}\frac{x_<}{x^8_>},
\end{equation}
where, $x_>$ and $x_<$ are the larger and smaller values of $x$ and
$x'$, respectively. Such a kernel smoothly decays to zero maintaining
the correct leading term in the asymptotic region, and giving the
correct long-range behavior for atom-diatom type interactions. For the
angular dimension $\theta$ a Taylor spline kernel is used:
\begin{equation}
 k^{[2]}(z,z') = 1 + z_<z_> + 2z^2_<z_> - \frac{2}{3}z^3_<,
\end{equation}
Here, $z_>$ and $z_<$ are the larger and smaller values of $z$ and
$z'$, respectively, and $z$ is defined as
\begin{equation}
z = \frac{1 - {\rm cos} \theta}{2},
\end{equation}
so that the values of $z \in [0,1]$.\\

\noindent
Finally, the 3-dimensional kernel is
\begin{equation}
\label{3dk}
 K({\bf{x}}, {\bf{x}}') = k^{[2,4]}(R,R')k^{[2,6]}(r,r')k^{[2]}(z,z'),
\end{equation}
where, ${\bf{x}}, {\bf{x}}'$ are $(R, r, z)$ and $(R', r', z')$,
respectively. The coefficients $\alpha_i$ and the RKHS representation
of the PES are evaluated by using a computationally efficient
toolkit.\cite{MM.rkhs:2017}.\\

\subsection{Bound State Calculations}
The bound states supported by the PESs considered in the present work
were calculated using the DVR3D suite of codes.\cite{dvr3d:2004}. The
nuclear time-independent Schr\"odinger equation for N$_3^+$ is solved
in a discrete variable representation (DVR) grid in Jacobi
coordinates.  The angular degree of freedom is defined by 56
Gauss-Legendre quadrature points while the radial degrees of freedom
are defined by Gauss-Laguerre quadratures, 72 points along $R$ and 48
points along $r$. The angular basis functions are expressed as
Legendre polynomials while the radial basis functions are constructed
using Morse oscillator functions with $r_e = 2.25$ a$_0$, $D_e = 0.32
E_{\rm h}$ and $\omega_e = 0.008 E_{\rm h}$ for $r$, and with $R_e =
3.8$ a$_0$, $D_e = 0.15 E_{\rm h}$ and $\omega_e = 0.0015 E_{\rm h}$
for $R$.  With these parameters the $r$ grid extends from 1.38 to 3.07
a$_0$ while the $R$ grid covers the range between 1.60 and 5.91
a$_0$. The $r_2$ embedding is used to calculate the rotationally
excited states, where the $z-$axis is parallel to $R$ in body-fixed
Jacobi coordinates. The vibrational wave functions are transformed
from Jacobi coordinates to symmetric $(R_{\rm N1N2}+R_{\rm
  N2N3})/\sqrt{2}$), asymmetric $(R_{\rm N1N2}-R_{\rm
  N2N3})/\sqrt{2}$) and bending coordinates using a Gaussian
kernel-based interpolation method where the terminal N atoms are N1
and N3 while the middle N atom is N2. The bending coordinate is
$\angle$ N1N2N3. Quantum numbers associated with the vibrational wave
functions are assigned by counting the nodal planes along each
coordinate.\\

\noindent
As an independent validation of the vibrational frequencies, time
dependent quantum mechanical (TDQM) calculations were carried out to
determine the bound states for $J = 0$. In this approach an
autocorrelation function is computed from the time propagation of an
initial wave packet followed by Fourier transformation to compute the
energy spectrum.\cite{dai96:3664,kon16:034303,konthesis} The initial
wave packet is a product of two Gaussian functions along the radial
$(R, r)$ coordinates and a function based on normalized associated
Legendre polynomials describes the angular coordinates.  The initial
wavepacket (WP) is located at $R = 3.6$ a$_0$ and $r = 2.1$ a$_0$. The
$R$ grid consists of 108 evenly spaced points from 1.55 to 9.575 a$_0$
while the $r$ grid has 70 points with a span from 1.1 to 4.55
a$_0$. The angular grid was defined by 56 Gauss-Legendre quadrature
points. The split operator method is used to compute the time
evolution of the WP.\cite{fei82:412} A sine damping function is
multiplied to the WP near the grid boundary to avoid unphysical
reflections from it. The autocorrelation function $A(t) =
\langle\psi_0|\psi_t\rangle$ is calculated at each time step. Finally
the eigen energy spectrum was obtained from the Fourier transform of
$A(t)$.  A window function (Normalized Hanning or Gaussian) is used to
reduce the noise in the spectra.\cite{dai95:1491,mah95:6057} The peak
positions are then determined by fitting a Gaussian function to each
peak.\\

\section{Results and Discussion}
\subsection{The Ground State Potential Energy Surface}
{\it Ab initio} energies up to 10 eV from the N$^+_3$ atomization
energy ($E_{\rm N}+E_{\rm N}+E_{\rm N^+}$) computed at the
CCSD(T)-F12b and MRCI+Q levels of theory are used as input for the
RKHS to construct the PESs.  First, the quality of the RKHS
representation was assessed.  Figure SI1 compares the {\it ab initio}
energies at the two different levels of theory with those obtained
from RKHS interpolations. The correlation coefficients ($R^2$) between
the reference calculations (CCSD(T)-F12b and MRCI+Q) and the RKHS
representation are $1 - 4\times10^{-7}$ and $1 - 3\times10^{-7}$,
respectively, for energies up to 6 eV, see Figure \ref{sifig:grid}.\\

\begin{figure}[htbp]
\includegraphics[width=0.85\textwidth]{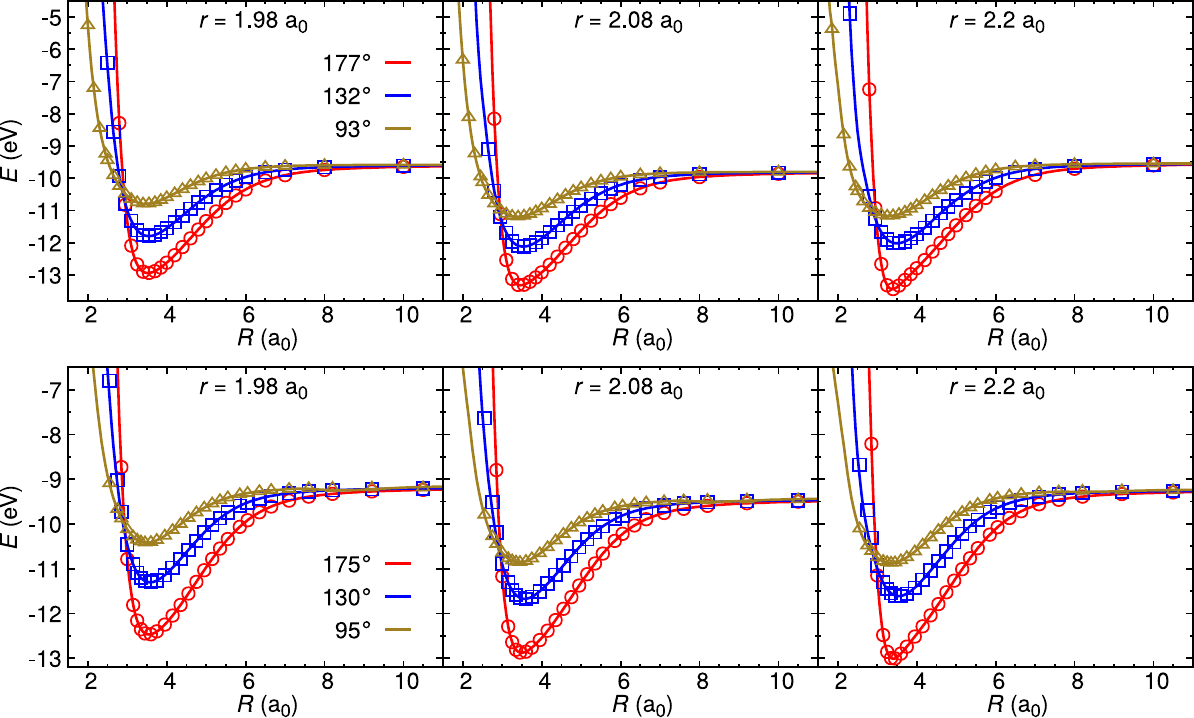}
\caption{Comparison between the {\it ab initio} and RKHS interpolated
  energies as a function of $R$. Upper panel represents the
  CCSD(T)-F12b PES and lower panel represents the MRCI+Q PES.}
\label{fig:comp1d}	
\end{figure}

\noindent
Furthermore, energies for 219 and 252 off-grid points were computed at
the CCSD(T)-F12b and MRCI+Q levels of theory, respectively, and the
corresponding energies from the 3d-RKHS PESs were evaluated. For the
CCSD(T)-F12b/aug-cc-pVTZ-f12 and MRCI+Q/aug-cc-pVTZ PESs the
respective correlation coefficients are 0.9999 and 0.9993 (Figure
\ref{sifig:offgrid}). Selected 1D cuts along $R$ for different values
of $r$ and $\theta$ are shown in Figure \ref{fig:comp1d}. These
comparisons establish the high quality of the representation of the
PES.\\

\begin{figure}[htbp]
\begin{center}
\includegraphics[width=0.85\textwidth]{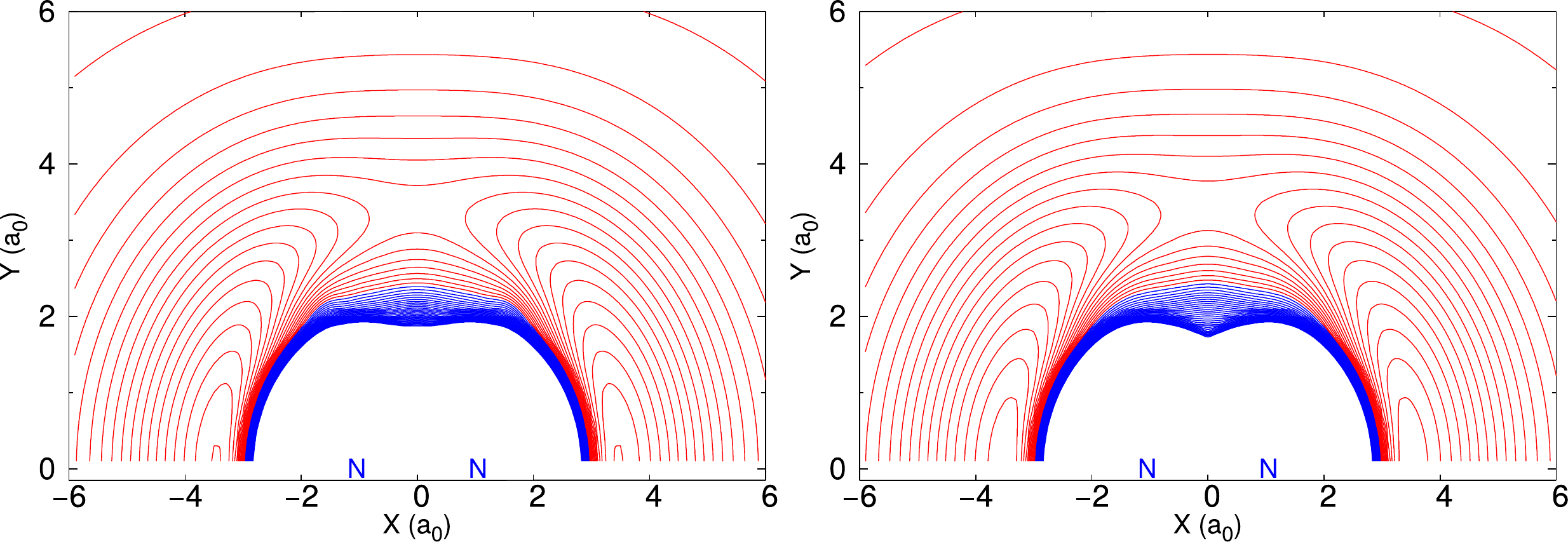}
\caption{Contour diagram for the analytical PESs (left: CCSD(T)-F12b
  PES and right: MRCI+Q PES) for $r = 2.080$ a$_0$.  Contour spacings
  are 0.2 eV. Blue and red lines represent positive and negative
  energies, respectively. The zero of energy is at the N$^+$($^3$P) +
  N$_2$(X$^1\Sigma_g^+$) asymptote.}
\label{fig:n3xy}
\end{center}
\end{figure}

\begin{figure}[htbp]
\begin{center}
\includegraphics[width=0.85\textwidth]{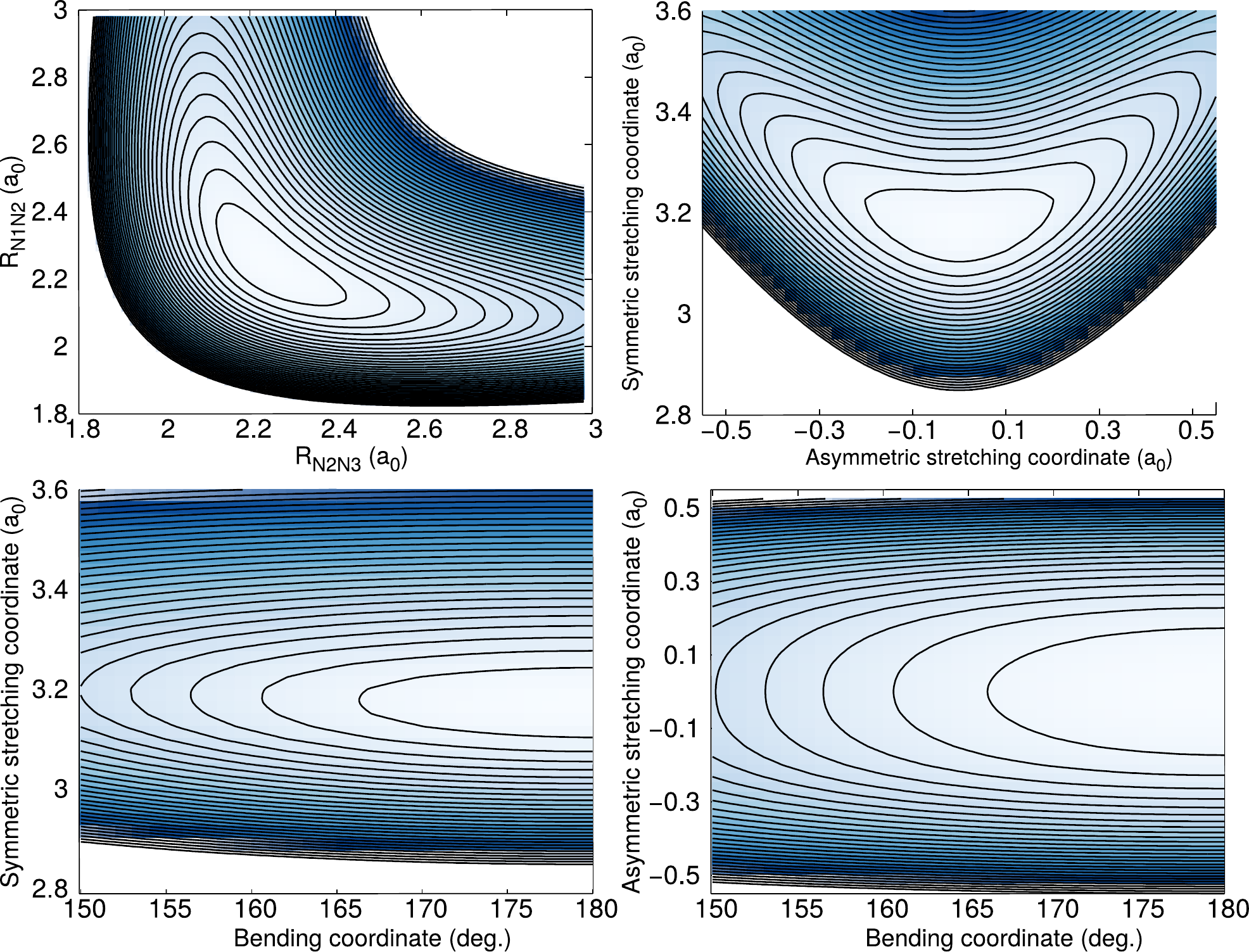}
\caption{Contour and color map representation of the the CCSD(T)-F12b
  RKHS PES. Upper left: In internal coordinates, $R_{\rm N1N2}$
  vs. $R_{\rm N2N3}$ with fixed $\angle ({\rm N1N2N3}) =
  180^\circ$. Upper right: In symmetric, asymmetric stretching,
  coordinates with fixed $\angle ({\rm N1N2N3}) = 180^\circ$. Lower
  left: In symmetric stretching and bending coordinates with a value
  of zero for the asymmetric stretching. Lower right: In asymmetric
  stretching and bending coordinates with a fixed value of 3.1725
  a$_0$ for the symmetric stretching coordinate. Spacing between the
  contours is 500 cm$^{-1}$}
\label{fig:sym}
\end{center}
\end{figure}

\noindent
A 2-d projection for an N1-N2 separation of 2.080 a$_0$ is shown in
Figure \ref{fig:n3xy} for both PESs. The linear configurations show a
deep minimum, which is the global minimum for the system. For the
CCSD(T)-F12b PES, the global minimum lies 3.649 eV below the
N$^+$+N$_2$ asymptote for $r_{\rm NN}$ = 2.243 a$_0$. This compares
well with the minimum obtained from electronic structure calculations
as 2.244 a$_0$. For the MRCI+Q PES the global minimum is 3.591 eV
below the N$^+$+N$_2$ asymptote for a linear symmetric structure with
$r_{\rm NN}$ = 2.256 a$_0$. Two-dimensional cuts in internal
coordinates are provided in Figure \ref{fig:sym} for the equilibrium
region which shows the symmetric nature of the PES. \\

\subsection{Electronic Structure of the Ground State PES}
\begin{figure}[htbp]
\begin{center}
\includegraphics[width=\textwidth]{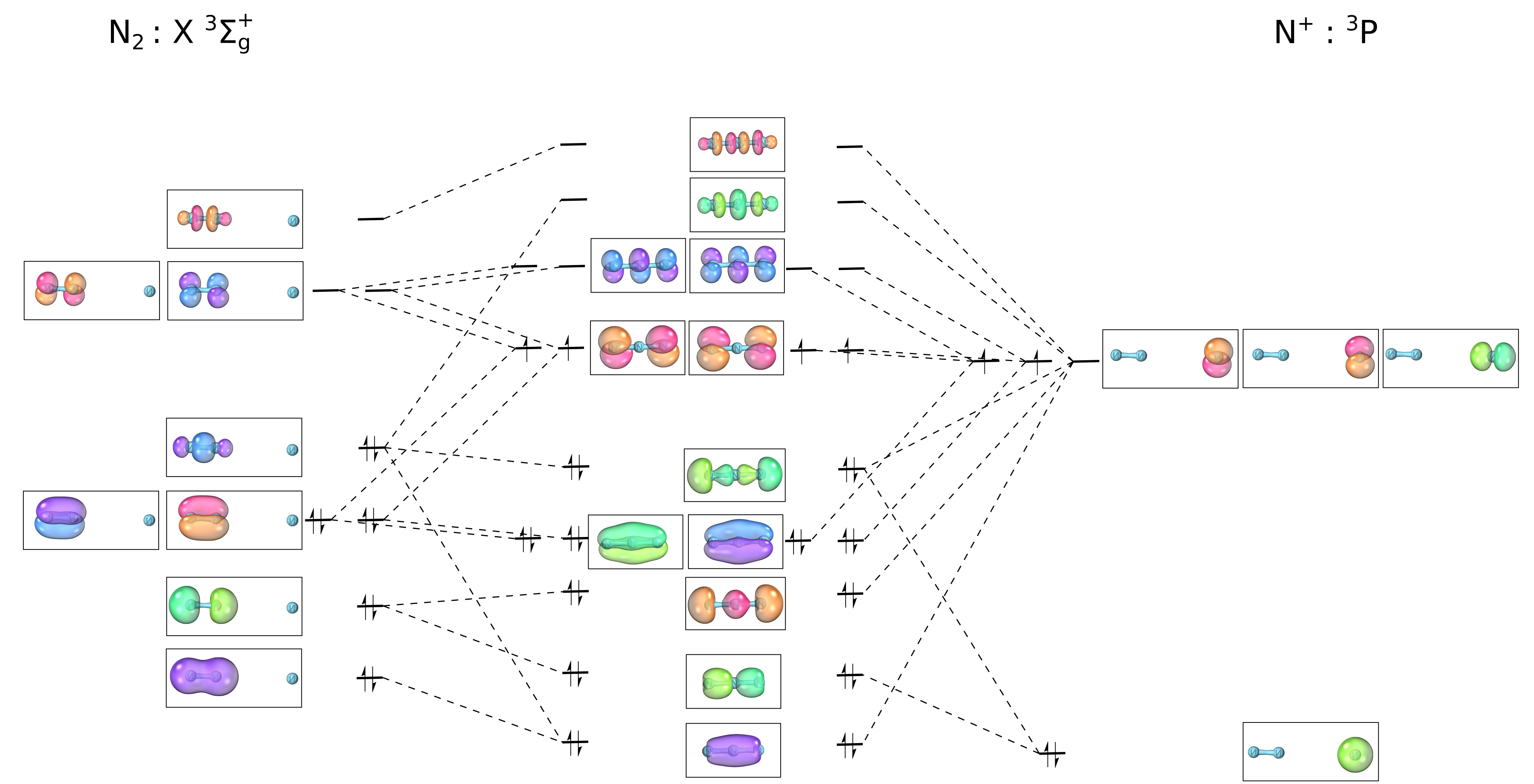}
\caption{MO diagram of the CASSCF valence space natural orbitals for linear
  bonding of N$^+$ to N$_2$. Single occupation of the degenerate
  $\pi_3$-nonbonding orbitals leads to a triplet ground state. Orbital
  visualization has been performed with
  IboView\cite{IboView}. The orbital isosurfaces enclose 80\% of the
  electron density.}
\label{fig:modiag_lin}
\end{center}    
\end{figure}

\begin{figure}[htbp]
\begin{center}
\includegraphics[width=\textwidth]{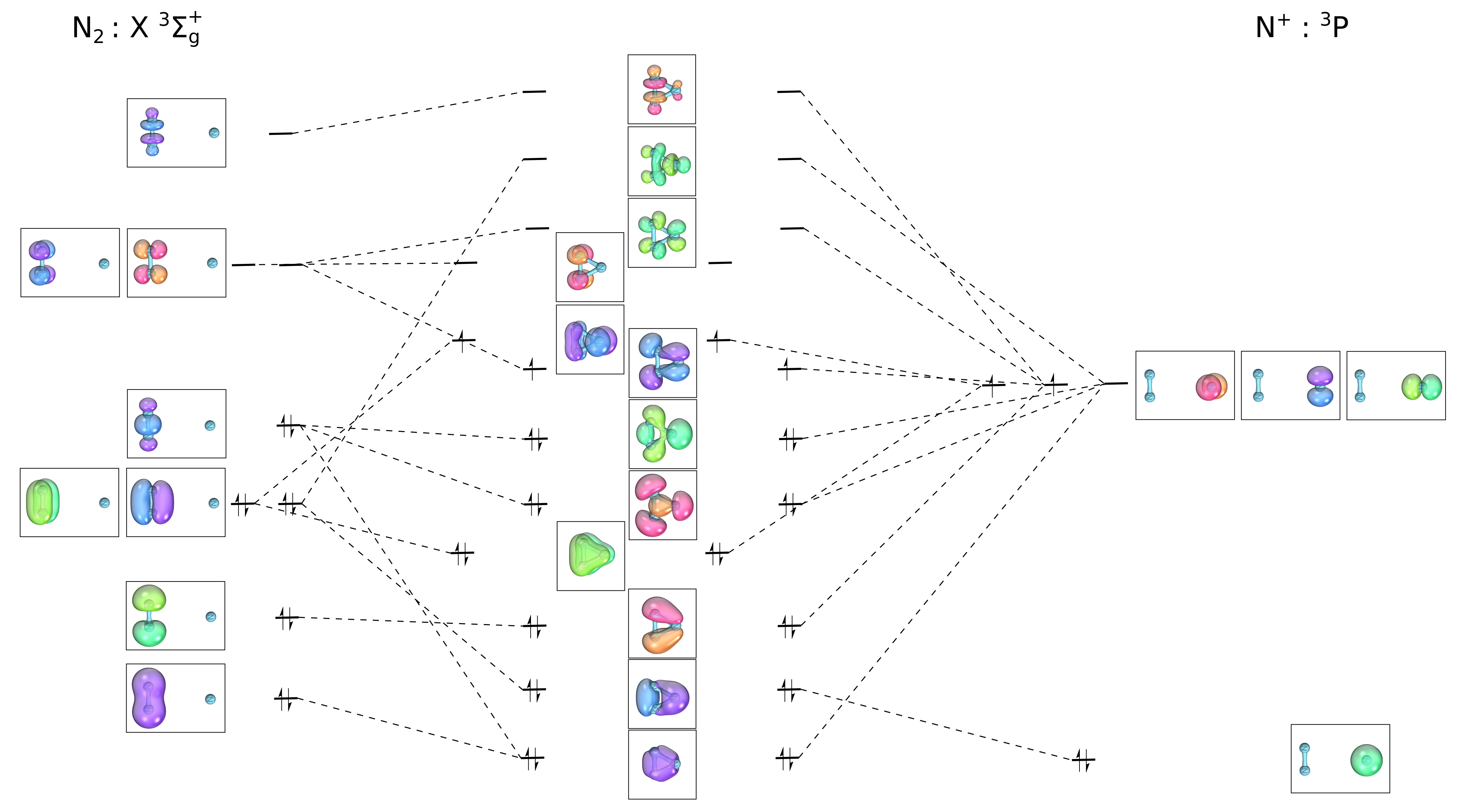}
\caption{MO diagram of the valence space natural orbitals for the
  T-shaped bonding of N$^+$ to N$_2$. The nitrogen $p$-orbitals
  recombine to form frontier MOs with different nodal structures that
  are close in energy but not degenerate.}
\label{fig:modiag_tshp}
\end{center}
\end{figure}

Molecular orbital (MO) diagrams for the bonding of N$^+$ to N$_2$
shown in Figure \ref{fig:modiag_lin} (``end-on'') and Figure
\ref{fig:modiag_tshp} (``side-on'') help explain the topography of the
PES in Figure \ref{fig:n3xy}. The MOs are obtained as CASSCF natural
orbitals of the valence space (i.\ e.\ the active space) and their
occupation is determined from the one or few dominant CASSCF CI
vectors. Ground state (X$^3 \Sigma_g^-$) linear N$^+_3$ has two
degenerate, singly occupied non-bonding $\pi_3$ orbitals. Since the
singly occupied orbitals of the triatomic complex smoothly transform
into degenerate $p$-orbitals on N$^+$($^3$P) upon linear dissociation,
a triplet ground state with one dominant configuration contributing to
the CASSCF wave function (normalized square norm of the CI coefficient
$> 65$\%) persists along the entire path. This characteristic also
explains the absence of a repulsive region at long range as a
consequence of an avoided crossing (see Figure \ref{fig:comp1d} at
$\theta=175^{\circ}$). A more detailed evolution of the corresponding
CASSCF natural orbitals is shown in Figure
\ref{sifig:modiag_lin_detail}. For covalent bonding distances $r
\approx R$ the natural orbitals are symmetry-adapted 3-center bonds
and their energy ordering is essentially determined by their $\pi-$ or
$\sigma-$ character and the number of nodal planes that partition the
bond axis.\\

\noindent
A triplet state ``side-on'' bonding of N$^+$ to N$_2$ also occurs
along a dissociation path without an avoided crossing of two
electronic states (with the absence of repulsive barrier in the 1-D
dissociation curve in Figure \ref{fig:comp1d} at $\theta=95^{\circ}$,
$130^{\circ}$) and one dominant triplet state electron
configuration. An MO diagram of the valence space CASSCF natural
orbitals is shown in Figure \ref{fig:modiag_tshp} with more details
reported in Figure \ref{sifig:modiag_tshp_detail}. The degeneracy of
the singly occupied orbitals breaks upon bending of the linear
structure. The two singly occupied orbitals in the triplet state
remain quasi-degenerate upon slight to moderate bending of the linear
molecule. Only for structures close to an equilateral triangle the
lowest singlet and triplet states are energetically close. The singlet
state is lower in energy than the triplet state only in a small region
of the PES, namely for $r\lesssim 2.3\ a_0$, $R\lesssim 2.6\ a_0$ and
$\theta \approx 90^\circ$. This triangular structure with
approximately equidistant atoms is, however, more than $\approx 2.8$
eV above the absolute minimum of the PES (see Figure \ref{fig:n3xy})
at the MRCI+Q level of theory. Hence, it can be assumed that the
excited singlet state's role on the molecular dynamics of the ground
state is minor.\\

\noindent
Remarkably, the transition from the linear to the triangular structure
on the triplet state evolves along one dominant CASSCF
configuration. This occurs because of a smooth transformation along
the reaction channel of the singly occupied $\pi_3$ non-bonding
orbitals in the linear structure to the singly occupied orbitals in
the T-shaped structure. Comparing these orbitals in Figures
\ref{fig:modiag_lin} and \ref{fig:modiag_tshp} indicate that the
planes of antisymmetry are conserved. Such a one-to-one mapping
emerges for all orbitals in the linear and T-shaped structure and
energy reordering of them along the reaction channel only occurs in
ways that do not qualitatively change the occupation number of a
natural orbital.\\

\noindent
It should be noted that the degeneracy of the three occupation
configurations of the $p$-orbitals in the $^3$P state of N$^+$ can be
interpreted as a multireference character of N$^+$. However, the
coupling of the three states in the Hamiltonian can always be removed
by appropriate orbital rotations even for the N$^+$--N$_2$ Van der
Waals complex. Therefore, a meaningful Hartree-Fock solution also
exists in these cases. In this context, it should be noted that the
difference in energy stemming from the difference in Hartree-Fock and
CASSCF occupied orbitals is strongly reduced by the subsequent coupled
cluster or MRCI computations, as the singles excitations in the
correlated methods effectively act as correcting orbital rotations
under the (higher-order) electron correlation effects.\\

\noindent
In summary, for an only mildly stretched N1-N2 covalent bond ($r <
2.7$ $a_0$), changes of $R$ and $\theta$ can be qualitatively
correctly described by a smooth evolution of a wave function with one
dominant triplet configuration. This explains why open-shell
unrestricted coupled cluster with a triplet state restricted
Hartree-Fock reference wave function performs very well in this region
of the PES. It is also interesting to note that the ground state
electronic structure of N$_3^+$ bears characteristics similar to those
of molecular oxygen, a very stable triplet ground state molecule.\\

\noindent
When the values of $r$ and $R$ both increase towards dissociation, the
multireference character of the systems quickly becomes
significant. Here only a full valence space CASSCF wave function can
be a reliable reference wave function. In the dissociation region two
energetically closely lying states emerge which are related by charge
transfer. Experimentally, N$^+$+N$_2$ is energetically favoured over
N+N$^+_2$ by 1.05 eV which is the difference in the first ionization
energies of N$_2$ (15.58 eV; Ref.\cite{trickl.n2:1989}) and that of
atomic nitrogen (14.53 eV; Ref.\cite{crchandbook:2007}). At the CASSCF
and MRCI+Q levels of theory, the computed ionization energies of N$_2$
are 16.33 eV and 15.52 eV, respectively, whereas those of atomic
nitrogen are 13.37 eV and 14.45 eV. Thus, MRCI+Q yields ionization
energies in good agreement with experiment whereas CASSCF does not.\\

\begin{figure}[htbp]
\begin{center}
\includegraphics[width=0.8\textwidth]{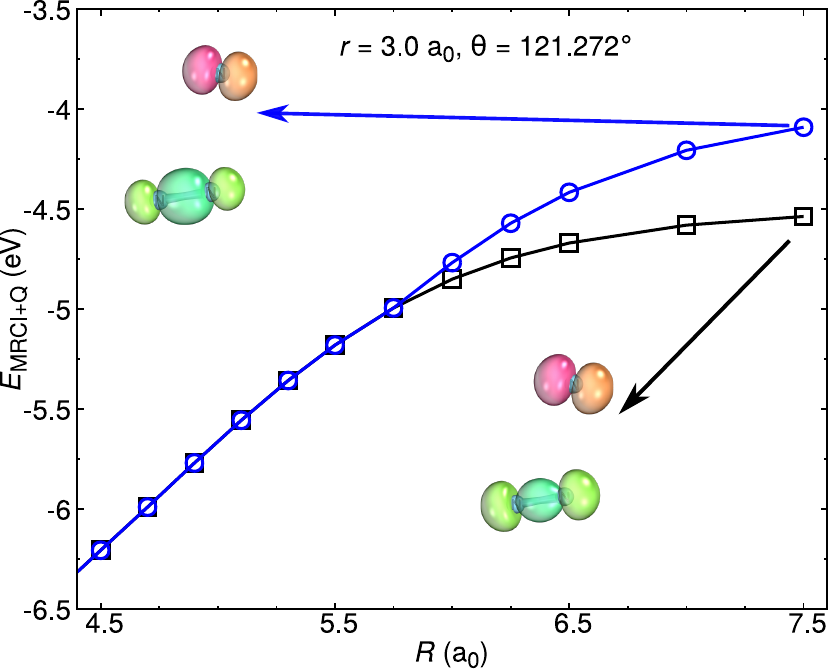}
\caption{1D cut of the {\it ab initio} PES along $R$ for
  $\theta=121.272 ^\circ$ and $r = 3.0$ a$_0$. Branching of the CASSCF
  solution that occurs when using initial guesses of orbitals that are
  connected to the N$^+$ + N$_2$ (blue) and N + N$_2^+$ (black)
  dissociation. Note the slight difference in spatial extent of the
  CASSCF natural orbitals (respective HOMOs of the two monomers, the
  density isosurface encloses 80\% of the electron density).}
\label{fig:splitting_casscf}
\end{center}
\end{figure}

\noindent
For stretched N$_2$ with $r = 3.0$ $a_0$ the charge transfer reaction
energy of N$^+$+N$_2$ $\to$ N+N$^+_2$ is 0.84 eV and --0.38 eV at the
CASSCF and MRCI+Q level of theory, respectively. Therefore, the
triplet ground state may become the charge transfer state associated
with the N+N$^+_2$ dissociation limit. Even though these are rather
high-energy regions of the PES, this aspect has to be monitored, as
the CASSCF algorithm will converge to different solutions depending on
the initial orbital guess if the natural orbitals between which the
charge transfer occurs are spatially sufficiently separated and can
therefore not mix. Figure \ref{fig:splitting_casscf} illustrates such
a case: at $r =3$ $a_0$ two different CASSCF solutions for the lowest
$^3$A$''$ state are obtained depending on the initial orbital guess
for values of $R$ above $\approx 6$ $a_0$. The natural orbital pairs
of the charge transfer are shown as inset and their slight difference
in spatial extent stems of the neutral or cationic character of the
moieties.\\

\noindent
In the region strong orbital mixing (formation of the covalent bond)
the two charge transfer states (N$^+$+N$_2$ and N+N$^+_2$) must mix
and therefore have to go through a conical intersection. The N$_3^+$
excited states that connect to this conical intersection for values of
$R$ smaller than the covalent bond limit ($R \lesssim 5.5$ a$_0$),
involve occupation of one of the antibonding orbitals of N$_3^+$ and
therefore steeply increase in energy for decreasing values of
$R$.\cite{rosmus.n3:1996} In fact, the MO diagrams in Figures
\ref{fig:modiag_lin} and \ref{fig:modiag_tshp} indicate that the
singly occupied orbitals of the N$_3^+$ ground state involve the $p_x$
and $p_y$ orbitals of of N$^+$. The orbital connections for collinear
N$^+$+N$_2$ in the MO diagram of Figure \ref{fig:modiag_lin} only
exist for orbitals with the same irreducible representation in
$C_{\infty v}$, the point group that is preserved along the collinear
reaction channel.  An electron count then yields that the $\pi$-system
of the N$_3^+$ ground state can only be connected to the $^3$P N$^+$
state with occupied $p_x$ and $p_y$ orbitals, as indicated.  The two
other of the three degenerate $^3$P states of N$^+$ then necessarily
connect to a high lying excited state of N$_3^+$ with an occupied
$\sigma^*$ orbital. This is also consistent with earlier
results.\cite{rosmus.n3:1996}\\

\noindent
In summary, for values of $R$ smaller than in the immediate region of
the covalent bond formation between the N$_2$ and N moieties all
(near-)degeneracies of the different triplet states are lifted and
restricted open shell Hartree-Fock is a meaningful reference wave
function. \\

\subsection{Vibrational Spectrum}
\begin{figure}[htbp]
\begin{center}
\includegraphics[width=\textwidth]{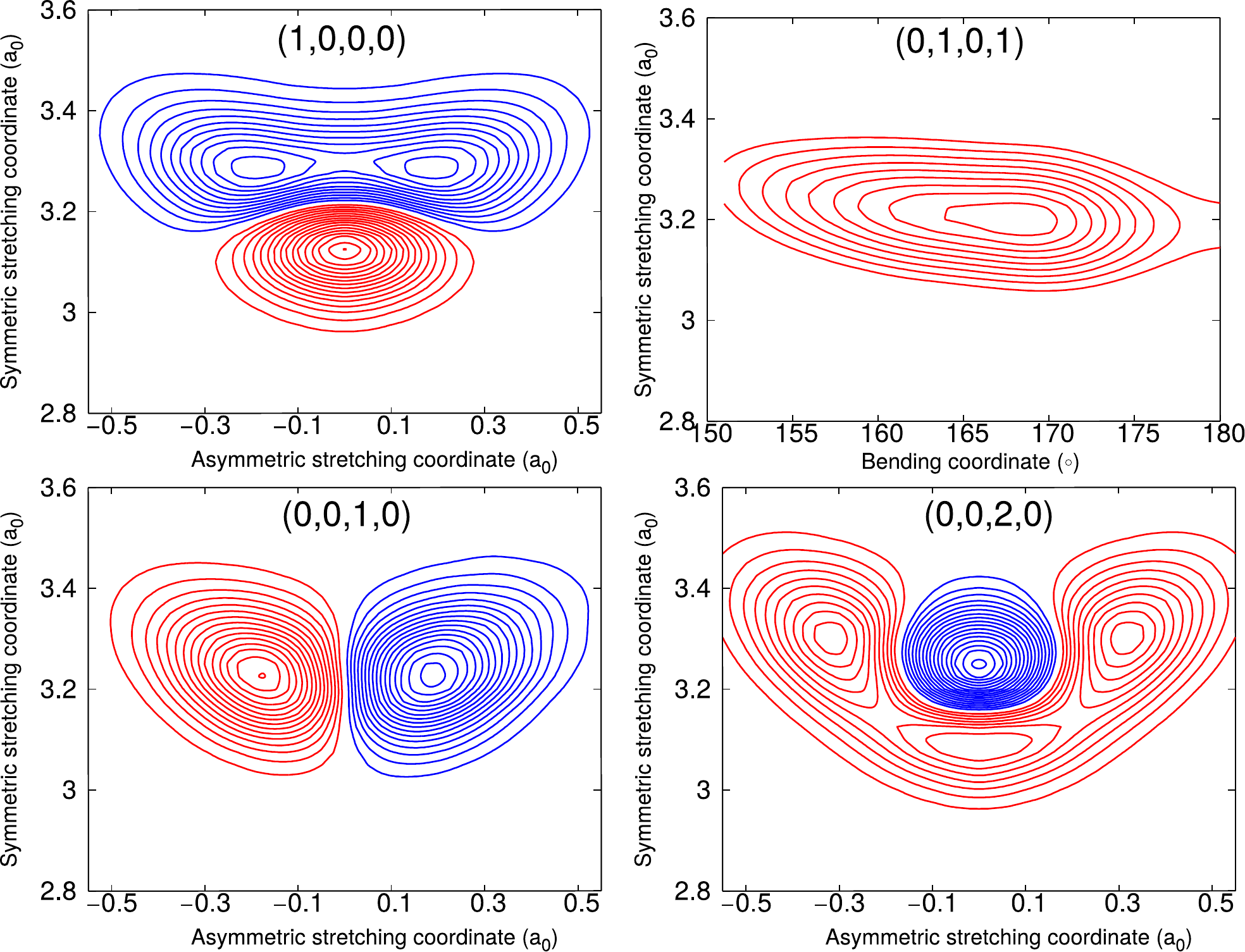}
\caption{Wavefunctions from DVR3D calculations for several low energy
  states, as indicated by the labels $(\nu_1 \nu_2 \nu_3 l)$.}
\label{fig:wave}
\end{center}
\end{figure}

The vibrational levels of the N$_{3}^{+}$ ion were obtained by solving
the nuclear Schr{\"o}dinger equation using the DVR3D\cite{dvr3d:2004}
suite of programs. The absolute and relative energies for the $J=0$
states on the CCSD(T)-F12 and MRCI+Q PESs using DVR3D and the solution
of the time dependent nuclear Schr\"odinger equation are given in
Table \ref{tab:tab1}. A few representative wavefunctions for the lower
levels are reported in Figure \ref{fig:wave} and additional ones are
given in Figures \ref{sifig:wave0200} to \ref{sifig:wave0400}.\\

\noindent
The results in Table \ref{tab:tab1} demonstrate that solutions from
the time-independent and time-dependent nuclear Schr\"odinger equation
are very close to one another - typically within less than 1
cm$^{-1}$. Hence, these energies can be considered converged. On the
other hand the absolute energies between the two PESs can differ by up
to 90 cm$^{-1}$. Given that a single reference treatment of the
electronic structure for the deeply bound region is a good
approximation it is expected that the results on the CCSD(T)-F12 PES
are more accurate than those on the MRCI+Q PES.\\

\begin{table}[ht!]
\caption{The 20 lowest bound states (in cm$^{-1}$) for N$^+_3$ with $J
  = 0$ obtained from DVR3D and TDQM calculations on the CCSD(T)-F12b
  and MRCI+Q PESs. The zero of energy is set at the minimum of the
  PES.}
    \begin{tabular}{lrrrr|lrrrr}
\hline
\hline
\multicolumn{5}{c}{CCSD(T)-F12b}&\multicolumn{5}{c}{MRCI+Q} \\
\hline
\multicolumn{3}{c}{Tot. Energy}&\multicolumn{2}{c}{Rel. Energy}&\multicolumn{3}{c}{Tot. Energy}&\multicolumn{2}{c}{Rel. Energy} \\
\hline
$n$ & DVR3D & TDQM & DVR3D & TDQM & $n$ & DVR3D & TDQM & DVR3D & TDQM \\
\hline
  1 & 1483.5  & 1483.6 &    0.0 &     0.0  &  1 & 1421.4 &  1421.4  &    0.0 &     0.0 \\
 2 & 2290.6  & 2289.9 &  807.0 &   806.3  &  2 & 2195.3 &  2195.3  &  773.9 &   773.9 \\
 3 & 2294.3  & 2295.6 &  810.8 &   811.9  &  3 & 2208.4 &  2208.5  &  787.1 &   787.1 \\
 4 & 2613.6  & 2613.4 & 1130.1 &  1129.8  &  4 & 2517.1 &  2517.2  & 1095.8 &  1095.8 \\
 5 & 3059.8  & 3059.8 & 1576.3 &  1576.2  &  5 & 2943.9 &  2944.0  & 1522.6 &  1522.6 \\
 6 & 3112.0  & 3112.1 & 1628.5 &  1628.4  &  6 & 3000.4 &  3000.5  & 1579.1 &  1579.1 \\
 7 & 3241.2  & 3241.2 & 1757.6 &  1757.5  &  7 & 3127.4 &  3127.5  & 1706.0 &  1706.1 \\
 8 & 3265.9  & 3265.9 & 1782.3 &  1782.3  &  8 & 3140.9 &  3141.0  & 1719.5 &  1719.6 \\
 9 & 3408.8  & 3408.7 & 1925.3 &  1925.1  &  9 & 3290.3 &  3290.4  & 1869.0 &  1869.0 \\
 10 & 3748.0  & 3748.2 & 2264.5 &  2264.6  & 10 & 3615.8 &  3615.9  & 2194.4 &  2194.5 \\
 11 & 3831.6  & 3831.6 & 2348.1 &  2348.0  & 11 & 3695.3 &  3695.3  & 2273.9 &  2273.9 \\
 12 & 3933.5  & 3933.6 & 2450.0 &  2449.9  & 12 & 3799.7 &  3799.7  & 2378.3 &  2378.3 \\
 13 & 4002.1  & 4002.2 & 2518.6 &  2518.6  & 13 & 3869.1 &  3869.2  & 2447.8 &  2447.8 \\
 14 & 4016.5  & 4016.5 & 2533.0 &  2532.9  & 14 & 3871.8 &  3872.0  & 2450.5 &  2450.6 \\
 15 & 4076.7  & 4076.6 & 2593.2 &  2593.0  & 15 & 3926.0 &  3926.1  & 2504.6 &  2504.7 \\
 16 & 4175.5  & 4176.2 & 2691.9 &  2692.5  & 16 & 4033.4 &  4033.5  & 2612.0 &  2612.1 \\
 17 & 4202.4  & 4202.4 & 2718.8 &  2718.7  & 17 & 4061.5 &  4061.5  & 2640.1 &  2640.2 \\
 18 & 4482.9  & 4483.0 & 2999.4 &  2999.4  & 18 & 4332.5 &  4332.6  & 2911.1 &  2911.2 \\
 19 & 4508.9  & 4509.0 & 3025.4 &  3025.3  & 19 & 4355.5 &  4355.6  & 2934.1 &  2934.2 \\
 20 & 4604.6  & 4604.7 & 3121.1 &  3121.1  & 20 & 4449.7 &  4449.8  & 3028.4 &  3028.4 \\
\hline
\hline
\end{tabular}
\label{tab:tab1}
\end{table}

\noindent
A comparison between the bound state energies and assignments from the
present and previous\cite{rosmus.n3:1994,jolibois:2009} work is given
in Table \ref{tab:compstates}. A first investigation of the lower
bound states was carried out on 3-dimensional PES
(``near-equilibrium'' PES\cite{rosmus.n3:1994}) based on internally
contracted CI calculations and a basis set similar to a Dunning cc-VQZ
basis set. This PES was fitted to a parametrized form and the bound
states were determined variationally.\cite{rosmus.n3:1994} In later
work,\cite{jolibois:2009} Born Oppenheimer dynamics were run at the
B3PW91/6-31G(d,p) level of theory at 283 K and 700 K and spectroscopic
features were extracted from the Fourier transform of the velocity
autocorrelation function.\\

\begin{table}[ht!]
 \caption{Lower bound states (in cm$^{-1}$) from the
   literature\cite{rosmus.n3:1994,jolibois:2009} and the present
   work. The assignment to harmonic quantum numbers $\nu_1$ (symmetric
   stretch), $\nu_2$ (bend), and $\nu_3$ (antisymmetric stretch) by
   node-counting is approximate due to strong couplings between the
   modes. The vibrational angular momentum quantum number is $l$.}
\begin{tabular}{l|rr|rr}
  \hline
  \hline
    $\nu_1 \nu_2 \nu_3 l$&  Ref.\cite{jolibois:2009}   &  Ref.\cite{rosmus.n3:1994} &  CCSD(T)-F12 & MRCI+Q\\
    \hline
0 1 0 1                          &  393    & 426  & 406 & 395  \\
0 0 1 0                          &  900    & 929  & 807 & 774  \\
0 2 0 0                          &  785    & 851  & 811 & 787  \\
1 0 0 0                          &  1040  & 1190 & 1130 & 1096 \\
0 1 1 1                          &  1238  & 1334 & 1193 & 1150 \\
0 3 0 1                          &  1173  & 1281 & 1220 & 1184 \\
1 1 0 1                          &  1402  & 1614 & 1529 & 1484 \\
0 2 1 0                          &  1681  & 1739 & 1576 & 1523 \\
0 4 0 0                          &  1561  & 1709 & 1629 & 1579 \\
0 0 2 0                          &  1795  & 1883 & 1758 & 1706 \\
1 0 1 0                          &  1905  & 1943 & 1782 & 1720 \\
1 2 0 0                          &  1991  & 2038 & 1925&  1869 \\
0 3 1 1                          &  2141  & 2148 & 1963 & 1899 \\
0 5 0 1                          &  2130  & 2141 & 2040 & 1979 \\
0 1 2 1                          &  2260  & 2281 & 2139 & 2078  \\
1 1 1 1                          &  2390  & 2344 & 2160 & 2088  \\
2 0 0 0                          &  2420  & 2396 & 2265 & 2194 \\ 
1 3 0 1                          &             & 2466 & 2324 & 2256  \\
\hline
\end{tabular}
\label{tab:compstates}
\end{table}

\noindent
It is observed that for the (100) symmetric stretch level (see Table
\ref{tab:compstates}) the present CCSD(T) calculations are consistent
with earlier calculations on a 3d-PES\cite{rosmus.n3:1994} and with
experiment but less so for the {\it ab initio} MD
simulations.\cite{jolibois:2009} Also, the MRCI+Q calculations find
the $\nu_1$ mode at 1096 cm$^{-1}$, 74 cm$^{-1}$ below the
experimental value (1170 cm$^{-1}$) and almost 100 cm$^{-1}$ below the
value obtained from the earlier bound state
calculations.\cite{rosmus.n3:1994} All methods agree that the (001)
antisymmetric stretch level lies below the (100) stretch which points
towards an unusual shape of the PES. For the different methods used
here this excitation is between 774 cm$^{-1}$ and 807 cm$^{-1}$ which
compares with frequencies at 900 cm$^{-1}$ or above from previous
work.\cite{rosmus.n3:1994,jolibois:2009} Finally, the (010) bending
vibration level is predicted to lie at around 400 cm$^{-1}$ by all
calculations. For these two fundamentals no experimental data are
available. The zero point energy from the present calculations is 1484
cm$^{-1}$ on the CCSD(T)-F12/aug-cc-pVTZ-f12 PES and 1421 cm$^{-1}$ on
the MRCI PES, compared with 1524 cm$^{-1}$ on the earlier
3d-PES.\cite{rosmus.n3:1994}\\

\noindent
The higher vibrational states (combination bands and overtones) and
their approximate assignments in terms of harmonic quantum numbers are
listed in Table \ref{tab:compstates}.  Assignments of quantum numbers
were made based on node counting but due to the strong couplings for
certain states, the labels for the quantum numbers are rather
approximate. As with the previous comparison between bound states on
the previous 3d-PES\cite{rosmus.n3:1994} and the
BO-dynamics\cite{jolibois:2009}, differences exist with the present
calculations. This is already expected based on the observations for
the fundamentals above, but also due to the anharmonic nature of many
vibrations and the strong couplings between them. It is found that the
bound states calculated on the MRCI+Q and CCSD(T)-F12 PESs from the
present work are consistent but can differ by several 10
cm$^{-1}$. Thus, the MRCI calculations, albeit somewhat lower in
accuracy than the CCSD(T)-F12 approach, provide a realistic
description of the energetics and couplings. Also, typically, the
bound states on these two PESs are lower than those on the previous
PES\cite{rosmus.n3:1994}, sometimes by up to 200 cm$^{-1}$. This may
be due to both the overall shape of the PES and the nature of the
anharmonic couplings. A few representative wavefunctions are reported
in Figure \ref{fig:wave}.\\

\noindent
Independently, the gas phase fundamentals for N$_{3}^{+}$ were also
obtained from the power spectra of the N-N distances using MD
simulations.\cite{salehi.n3m:2019} For this, $NVE$ simulations were
run using CHARMM\cite{charmm:2009} with the RKHS-interpolated
CCSD(T)-F12b PES at a temperature of 10 K. The $\nu_1$, $\nu_2$, and
$\nu_3$ modes of N$_{3}^{+}$ are found to lie at 1275 cm$^{-1}$, 357
cm$^{-1}$ and 917 cm$^{-1}$, respectively, consistent with the DVR3D
results, see Table \ref{tab:compstates}.\\

\section{Conclusions}
The current work focuses on the three dimensional PES and the
low-lying vibrations of the N$_3^+$ radical cation in its $^3
\Sigma_{\rm g}^-$ ground state through electronic structure and
rigorous quantum bound state calculations. Ab initio energies at the
CCSD(T)-F12b and MRCI+Q level of theory were represented as a
reproducing kernel Hilbert space (RKHS) to generate PESs for the
electronic ground state of the ion. A detailed analysis of the
electronic structure of the ground state PES was performed to assess
the validity of Hartree-Fock as a reference wave function for coupled
cluster and to rationalize the topology of the PESs. The
multireference character of the electronic structure may vary from a)
``absent'' around the global minimum of the PES to b) ``weak'' along
the dissociation paths with one covalent N-N bond only mildly
stretched, and c) ``strong'' when two covalent bonds are stretched far
from their equilibrium distance at the same time. For this latter case
the role of energetically low lying charge transfer states was also
investigated.\\

\noindent
The vibrational bound states supported by each PES are calculated by
solving the time-independent and time-dependent nuclear Schr\"odinger
equation. For the symmetric stretching frequency, the only one
available from experiment, quite good agreement is found with the
results on the CCSD(T)-F12 PES. Further verification of the calculated
vibrational frequencies is provided through calculations on an MRCI+Q
PES and power spectra determined from MD simulations. In summary, it
is found that single-reference open-shell coupled cluster theory is
able to yield a reliable high-quality PES for the vibrational bound
states of the electronic ground state of N$_3^+$.\\

\noindent
This work demonstrates that MRCI+Q calculations provide a realistic
description of the energetics of this challenging system by direct
comparison with vibrational states from CCSD(T)-F12 calculations, both
methods using an aug-cc-pVTZ basis set. As the higher electronically
excited states require explicit inclusion of multi reference effects,
the present work also lays the groundwork for future exploration of
the reactive dynamics of the N$_3^+$ ion, similar to previous studies
of other N-containing neutrals, such as C+NO\cite{MM.cno:2018} or
N+NO\cite{MM.no2:2020}. For this, further studies of electronically
excited states of N$_3^+$ are required.\\

\section*{Acknowledgment}
Support by the Swiss National Science Foundation through grants
200021-117810, the NCCR MUST (to MM), and the University of Basel is
also acknowledged. Part of this work was supported by the United State
Department of the Air Force, which is gratefully acknowledged (to
MM). This work was supported by the Australian Research Council
Discovery Project Grants (DP150101427, DP160100474). The authors
acknowledge fruitful discussions with Prof.\ Stefan Willitsch. MM
acknowledges the Department of Chemistry of Melbourne University for a
Wilsmore Fellowship during which this work has been initiated.\\

\begin{figure}[htbp]
\begin{center}
\includegraphics[scale=0.5]{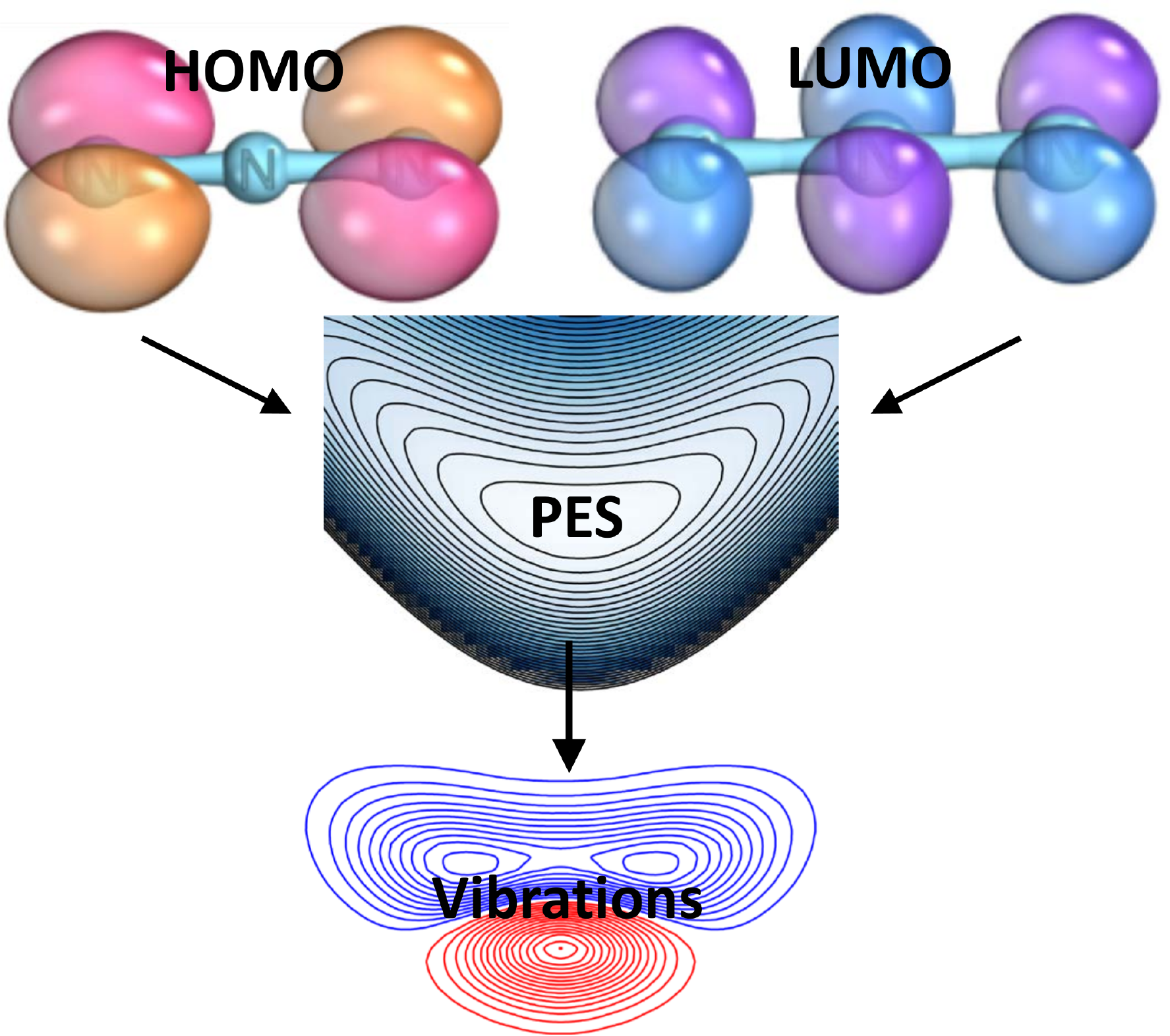}
\caption{TOC graphics}
\end{center}
\end{figure}

\bibliography{n3p}

\providecommand{\latin}[1]{#1}
\makeatletter
\providecommand{\doi}
  {\begingroup\let\do\@makeother\dospecials
  \catcode`\{=1 \catcode`\}=2 \doi@aux}
\providecommand{\doi@aux}[1]{\endgroup\texttt{#1}}
\makeatother
\providecommand*\mcitethebibliography{\thebibliography}
\csname @ifundefined\endcsname{endmcitethebibliography}
  {\let\endmcitethebibliography\endthebibliography}{}
\begin{mcitethebibliography}{40}
\providecommand*\natexlab[1]{#1}
\providecommand*\mciteSetBstSublistMode[1]{}
\providecommand*\mciteSetBstMaxWidthForm[2]{}
\providecommand*\mciteBstWouldAddEndPuncttrue
  {\def\EndOfBibitem{\unskip.}}
\providecommand*\mciteBstWouldAddEndPunctfalse
  {\let\EndOfBibitem\relax}
\providecommand*\mciteSetBstMidEndSepPunct[3]{}
\providecommand*\mciteSetBstSublistLabelBeginEnd[3]{}
\providecommand*\EndOfBibitem{}
\mciteSetBstSublistMode{f}
\mciteSetBstMaxWidthForm{subitem}{(\alph{mcitesubitemcount})}
\mciteSetBstSublistLabelBeginEnd
  {\mcitemaxwidthsubitemform\space}
  {\relax}
  {\relax}

\bibitem[de~Petris \latin{et~al.}(2006)de~Petris, Cartoni, Angelini, Ursini,
  Bottoni, and Calvaresi]{calvaresi:2006}
de~Petris,~G.; Cartoni,~A.; Angelini,~G.; Ursini,~O.; Bottoni,~A.;
  Calvaresi,~M. The N$_{3}^{+}$ reactivity in ionized gases containing sulfur,
  nitrogen, and carbon oxides. \emph{ChemPhysChem} \textbf{2006}, \emph{7},
  2105--2114\relax
\mciteBstWouldAddEndPuncttrue
\mciteSetBstMidEndSepPunct{\mcitedefaultmidpunct}
{\mcitedefaultendpunct}{\mcitedefaultseppunct}\relax
\EndOfBibitem
\bibitem[Anicich \latin{et~al.}(2000)Anicich, Milligan, Fairley, and
  McEwan]{mcewan:2000}
Anicich,~V.; Milligan,~D.; Fairley,~D.; McEwan,~M. Thermolecular ion-molecule
  reactions in Titan's atmosphere, I - Principal ions with principal neutrals.
  \emph{Icarus} \textbf{2000}, \emph{146}, 118--124\relax
\mciteBstWouldAddEndPuncttrue
\mciteSetBstMidEndSepPunct{\mcitedefaultmidpunct}
{\mcitedefaultendpunct}{\mcitedefaultseppunct}\relax
\EndOfBibitem
\bibitem[Bennett \latin{et~al.}(1996)Bennett, Maier, Chambaud, and
  Rosmus]{rosmus.n3:1996}
Bennett,~F.~R.; Maier,~J.~P.; Chambaud,~G.; Rosmus,~P. Photodissociation,
  charge and atom transfer processes in electronically excited states of
  N$_{3}^{+}$. \emph{Chem. Phys.} \textbf{1996}, \emph{209}, 275 -- 280\relax
\mciteBstWouldAddEndPuncttrue
\mciteSetBstMidEndSepPunct{\mcitedefaultmidpunct}
{\mcitedefaultendpunct}{\mcitedefaultseppunct}\relax
\EndOfBibitem
\bibitem[Midey \latin{et~al.}({2004})Midey, Miller, and
  Viggiano]{viggiano.n3p:2004}
Midey,~A.; Miller,~T.; Viggiano,~A. {Reactions of N$^+$, N$_2^{+}$, and
  N$_3^{+}$ with NO from 300 to 1400 K}. \emph{J. Chem. Phys.} \textbf{{2004}},
  \emph{{121}}, {6822--6829}\relax
\mciteBstWouldAddEndPuncttrue
\mciteSetBstMidEndSepPunct{\mcitedefaultmidpunct}
{\mcitedefaultendpunct}{\mcitedefaultseppunct}\relax
\EndOfBibitem
\bibitem[Friedmann \latin{et~al.}({1994})Friedmann, Soliva, Nizkorodov, Bieske,
  and Maier]{maier.n3p:1994}
Friedmann,~A.; Soliva,~A.; Nizkorodov,~S.; Bieske,~E.; Maier,~J. {A$^{3} \Pi
  _{u}$ -- X$^{3} \Sigma_{g}^{-}$ electronic spectrum of N$_{3}^{+}$}. \emph{J.
  Phys. Chem.} \textbf{{1994}}, \emph{{98}}, {8896--8902}\relax
\mciteBstWouldAddEndPuncttrue
\mciteSetBstMidEndSepPunct{\mcitedefaultmidpunct}
{\mcitedefaultendpunct}{\mcitedefaultseppunct}\relax
\EndOfBibitem
\bibitem[Dyke \latin{et~al.}(1982)Dyke, Jonathan, Lewis, and Morris]{dyke:1982}
Dyke,~J.; Jonathan,~N.; Lewis,~A.; Morris,~A. Vacuum ultraviolet photoelectron
  spectroscopy of transient species. \emph{Mol. Phys.} \textbf{1982},
  \emph{47}, 1231--1240\relax
\mciteBstWouldAddEndPuncttrue
\mciteSetBstMidEndSepPunct{\mcitedefaultmidpunct}
{\mcitedefaultendpunct}{\mcitedefaultseppunct}\relax
\EndOfBibitem
\bibitem[Archibald and Sabin(1971)Archibald, and Sabin]{archibald:1971}
Archibald,~T.~W.; Sabin,~J.~R. Theoretical investigation of the electronic
  structure and properties of N$_{3}^{-}$, N$_3$, and N$_3^+$. \emph{J. Chem.
  Phys.} \textbf{1971}, \emph{55}, 1821--1829\relax
\mciteBstWouldAddEndPuncttrue
\mciteSetBstMidEndSepPunct{\mcitedefaultmidpunct}
{\mcitedefaultendpunct}{\mcitedefaultseppunct}\relax
\EndOfBibitem
\bibitem[Cai \latin{et~al.}(1992)Cai, Wang, and Xiao]{cai:1992}
Cai,~Z.-L.; Wang,~Y.-F.; Xiao,~H.-M. Ab initio study of low-lying electronic
  states of the N$_3^+$ ion. \emph{Chem. Phys.} \textbf{1992}, \emph{164}, 377
  -- 381\relax
\mciteBstWouldAddEndPuncttrue
\mciteSetBstMidEndSepPunct{\mcitedefaultmidpunct}
{\mcitedefaultendpunct}{\mcitedefaultseppunct}\relax
\EndOfBibitem
\bibitem[Tian \latin{et~al.}({1988})Tian, Facelli, and Michl]{tian:1988}
Tian,~R.; Facelli,~J.; Michl,~J. {Vibrational and electronic spectra of
  matrix-isolated N$_{3}$ and N$_{3}^{-}$}. \emph{J. Phys. Chem.}
  \textbf{{1988}}, \emph{{92}}, {4073--4079}\relax
\mciteBstWouldAddEndPuncttrue
\mciteSetBstMidEndSepPunct{\mcitedefaultmidpunct}
{\mcitedefaultendpunct}{\mcitedefaultseppunct}\relax
\EndOfBibitem
\bibitem[Chambaud \latin{et~al.}({1994})Chambaud, Rosmus, Bennett, Maier, and
  Spielfiedel]{rosmus.n3:1994}
Chambaud,~G.; Rosmus,~P.; Bennett,~F.; Maier,~J.; Spielfiedel,~A. {Vibrational
  motion in the X$^{3} \Sigma{_{g}}^{-}$- state of N$_{3}^{+}$}. \emph{Chem.
  Phys. Lett.} \textbf{{1994}}, \emph{{231}}, {9--12}\relax
\mciteBstWouldAddEndPuncttrue
\mciteSetBstMidEndSepPunct{\mcitedefaultmidpunct}
{\mcitedefaultendpunct}{\mcitedefaultseppunct}\relax
\EndOfBibitem
\bibitem[Jolibois \latin{et~al.}({2009})Jolibois, Maron, and
  Ramirez-Solis]{jolibois:2009}
Jolibois,~F.; Maron,~L.; Ramirez-Solis,~A. {Ab initio molecular dynamics
  studies on the lowest triplet and singlet potential surfaces of the azide
  cation: Anharmonic effects on the vibrational spectra of linear and cyclic
  N$_{3}^{+}$}. \emph{{J. Mol. Struc.-THEOCHEM}} \textbf{{2009}}, \emph{{899}},
  {9--17}\relax
\mciteBstWouldAddEndPuncttrue
\mciteSetBstMidEndSepPunct{\mcitedefaultmidpunct}
{\mcitedefaultendpunct}{\mcitedefaultseppunct}\relax
\EndOfBibitem
\bibitem[Werner \latin{et~al.}(2012)Werner, Knowles, Knizia, Manby, and
  Sch{\"u}tz]{molpro}
Werner,~H.-J.; Knowles,~P.~J.; Knizia,~G.; Manby,~F.~R.; Sch{\"u}tz,~M. Molpro:
  A general-purpose quantum chemistry program package. \emph{WIREs Comput. Mol.
  Sci.} \textbf{2012}, \emph{2}, 242--253\relax
\mciteBstWouldAddEndPuncttrue
\mciteSetBstMidEndSepPunct{\mcitedefaultmidpunct}
{\mcitedefaultendpunct}{\mcitedefaultseppunct}\relax
\EndOfBibitem
\bibitem[Werner \latin{et~al.}(2019)Werner, Knowles, Knizia, Manby, and
  et~al.]{MOLPRO_brief}
Werner,~H.-J.; Knowles,~P.~J.; Knizia,~G.; Manby,~F.~R.; et~al.,~M.~S. MOLPRO,
  version 2019.1, A package of ab initio programs. 2019\relax
\mciteBstWouldAddEndPuncttrue
\mciteSetBstMidEndSepPunct{\mcitedefaultmidpunct}
{\mcitedefaultendpunct}{\mcitedefaultseppunct}\relax
\EndOfBibitem
\bibitem[Knizia \latin{et~al.}(2009)Knizia, Adler, and Werner]{knizia:2009}
Knizia,~G.; Adler,~T.~B.; Werner,~H.-J. {Simplified CCSD(T)-F12 methods: Theory
  and benchmarks}. \emph{J. Chem. Phys.} \textbf{2009}, \emph{130},
  054104\relax
\mciteBstWouldAddEndPuncttrue
\mciteSetBstMidEndSepPunct{\mcitedefaultmidpunct}
{\mcitedefaultendpunct}{\mcitedefaultseppunct}\relax
\EndOfBibitem
\bibitem[Peterson \latin{et~al.}(2008)Peterson, Adler, and
  Werner]{Kirk:VnZF1208}
Peterson,~K.~A.; Adler,~T.~B.; Werner,~H.-J. {Systematically convergent basis
  sets for explicitly correlated wavefunctions: The atoms H, He, B-Ne and
  Al-Ar}. \emph{J. Chem. Phys.} \textbf{2008}, \emph{128}, 084102\relax
\mciteBstWouldAddEndPuncttrue
\mciteSetBstMidEndSepPunct{\mcitedefaultmidpunct}
{\mcitedefaultendpunct}{\mcitedefaultseppunct}\relax
\EndOfBibitem
\bibitem[Werner and Knowles(1988)Werner, and Knowles]{wer88:5803}
Werner,~H.-J.; Knowles,~P.~J. An efficient internally contracted
  multiconfiguration\-reference configuration interaction method. \emph{J.
  Chem. Phys.} \textbf{1988}, \emph{89}, 5803--5814\relax
\mciteBstWouldAddEndPuncttrue
\mciteSetBstMidEndSepPunct{\mcitedefaultmidpunct}
{\mcitedefaultendpunct}{\mcitedefaultseppunct}\relax
\EndOfBibitem
\bibitem[Knowles and Werner(1988)Knowles, and Werner]{kno88:514}
Knowles,~P.~J.; Werner,~H.-J. An efficient method for the evaluation of
  coupling coefficients in configuration interaction calculations. \emph{Chem.
  Phys. Lett.} \textbf{1988}, \emph{145}, 514 -- 522\relax
\mciteBstWouldAddEndPuncttrue
\mciteSetBstMidEndSepPunct{\mcitedefaultmidpunct}
{\mcitedefaultendpunct}{\mcitedefaultseppunct}\relax
\EndOfBibitem
\bibitem[Dunning(1989)]{dun89:1007}
Dunning,~T.~H. Gaussian basis sets for use in correlated molecular
  calculations. I. The atoms boron through neon and hydrogen. \emph{J. Chem.
  Phys.} \textbf{1989}, \emph{90}, 1007--1023\relax
\mciteBstWouldAddEndPuncttrue
\mciteSetBstMidEndSepPunct{\mcitedefaultmidpunct}
{\mcitedefaultendpunct}{\mcitedefaultseppunct}\relax
\EndOfBibitem
\bibitem[Langhoff and Davidson(1974)Langhoff, and Davidson]{davidson:1974}
Langhoff,~S.; Davidson,~E. Configuration interaction calculations on nitrogen
  molecule. \emph{Int. J. Quant. Chem.} \textbf{1974}, \emph{8}, 61--72\relax
\mciteBstWouldAddEndPuncttrue
\mciteSetBstMidEndSepPunct{\mcitedefaultmidpunct}
{\mcitedefaultendpunct}{\mcitedefaultseppunct}\relax
\EndOfBibitem
\bibitem[Werner and Knowles(1985)Werner, and Knowles]{wen85:5053}
Werner,~H.-J.; Knowles,~P.~J. A second order multiconfiguration SCF procedure
  with optimum convergence. \emph{J. Chem. Phys.} \textbf{1985}, \emph{82},
  5053--5063\relax
\mciteBstWouldAddEndPuncttrue
\mciteSetBstMidEndSepPunct{\mcitedefaultmidpunct}
{\mcitedefaultendpunct}{\mcitedefaultseppunct}\relax
\EndOfBibitem
\bibitem[Knowles and Werner(1985)Knowles, and Werner]{kno85:259}
Knowles,~P.~J.; Werner,~H.-J. An efficient second-order MC SCF method for long
  configuration expansions. \emph{Chem. Phys. Lett.} \textbf{1985}, \emph{115},
  259 -- 267\relax
\mciteBstWouldAddEndPuncttrue
\mciteSetBstMidEndSepPunct{\mcitedefaultmidpunct}
{\mcitedefaultendpunct}{\mcitedefaultseppunct}\relax
\EndOfBibitem
\bibitem[Werner and Meyer(1980)Werner, and Meyer]{wer80:2342}
Werner,~H.-J.; Meyer,~W. A quadratically convergent
  multiconfiguration-selfconsistent field method with simultaneous optimization
  of orbitals and CI coefficients. \emph{J. Chem. Phys.} \textbf{1980},
  \emph{73}, 2342--2356\relax
\mciteBstWouldAddEndPuncttrue
\mciteSetBstMidEndSepPunct{\mcitedefaultmidpunct}
{\mcitedefaultendpunct}{\mcitedefaultseppunct}\relax
\EndOfBibitem
\bibitem[Kreplin \latin{et~al.}(2019)Kreplin, Knowles, and Werner]{werner:2019}
Kreplin,~D.~A.; Knowles,~P.~J.; Werner,~H.-J. Second-order MCSCF optimization
  revisited. I. Improved algorithms for fast and robust second-order CASSCF
  convergence. \emph{J. Chem. Phys.} \textbf{2019}, \emph{150}\relax
\mciteBstWouldAddEndPuncttrue
\mciteSetBstMidEndSepPunct{\mcitedefaultmidpunct}
{\mcitedefaultendpunct}{\mcitedefaultseppunct}\relax
\EndOfBibitem
\bibitem[Ho and Rabitz(1996)Ho, and Rabitz]{rabitz:1996}
Ho,~T.-S.; Rabitz,~H. A general method for constructing multidimensional
  molecular potential energy surfaces from ab initio calculations. \emph{J.
  Chem. Phys.} \textbf{1996}, \emph{104}, 2584--2597\relax
\mciteBstWouldAddEndPuncttrue
\mciteSetBstMidEndSepPunct{\mcitedefaultmidpunct}
{\mcitedefaultendpunct}{\mcitedefaultseppunct}\relax
\EndOfBibitem
\bibitem[Unke and Meuwly(2017)Unke, and Meuwly]{MM.rkhs:2017}
Unke,~O.~T.; Meuwly,~M. Toolkit for the construction of Reproducing
  Kernel-Based representations of data: Application to multidimensional
  potential energy surfaces. \emph{J. Chem. Inf. Model.} \textbf{2017},
  \emph{57}, 1923--1931\relax
\mciteBstWouldAddEndPuncttrue
\mciteSetBstMidEndSepPunct{\mcitedefaultmidpunct}
{\mcitedefaultendpunct}{\mcitedefaultseppunct}\relax
\EndOfBibitem
\bibitem[Tennyson \latin{et~al.}(2004)Tennyson, Kostin, Barletta, Harris,
  Polyansky, Ramanlal, and Zobov]{dvr3d:2004}
Tennyson,~J.; Kostin,~M.~A.; Barletta,~P.; Harris,~G.~J.; Polyansky,~O.~L.;
  Ramanlal,~J.; Zobov,~N.~F. DVR3D: a program suite for the calculation of
  rotation-vibration spectra of triatomic molecules. \emph{Comput. Phys.
  Commun.} \textbf{2004}, \emph{163}, 85 -- 116\relax
\mciteBstWouldAddEndPuncttrue
\mciteSetBstMidEndSepPunct{\mcitedefaultmidpunct}
{\mcitedefaultendpunct}{\mcitedefaultseppunct}\relax
\EndOfBibitem
\bibitem[Dai and Zhang(1996)Dai, and Zhang]{dai96:3664}
Dai,~J.; Zhang,~J. Z.~H. {Time-dependent spectral calculation of bound and
  resonance energies of HO$_2$}. \emph{J. Chem. Phys.} \textbf{1996},
  \emph{104}, 3664--3671\relax
\mciteBstWouldAddEndPuncttrue
\mciteSetBstMidEndSepPunct{\mcitedefaultmidpunct}
{\mcitedefaultendpunct}{\mcitedefaultseppunct}\relax
\EndOfBibitem
\bibitem[Koner \latin{et~al.}(2016)Koner, Barrios, Gonz\'{a}lez-Lezana, and
  Panda]{kon16:034303}
Koner,~D.; Barrios,~L.; Gonz\'{a}lez-Lezana,~T.; Panda,~A.~N. {Scattering study
  of the Ne + NeH$^+$($v_0 = 0, j_0 = 0)$ $\rightarrow$ NeH$^+$ + Ne reaction
  on an ab initio based analytical potential energy surface}. \emph{J. Chem.
  Phys.} \textbf{2016}, \emph{144}, 034303\relax
\mciteBstWouldAddEndPuncttrue
\mciteSetBstMidEndSepPunct{\mcitedefaultmidpunct}
{\mcitedefaultendpunct}{\mcitedefaultseppunct}\relax
\EndOfBibitem
\bibitem[Koner(2016)]{konthesis}
Koner,~D. \emph{{Scattering studies of proton transfer reactions between rare
  gas atoms}}; Indian Institute of Technology Guwahati, 2016\relax
\mciteBstWouldAddEndPuncttrue
\mciteSetBstMidEndSepPunct{\mcitedefaultmidpunct}
{\mcitedefaultendpunct}{\mcitedefaultseppunct}\relax
\EndOfBibitem
\bibitem[Feit \latin{et~al.}(1982)Feit, J.~A.~Fleck, and Steiger]{fei82:412}
Feit,~M.~D.; J.~A.~Fleck,~J.; Steiger,~A. {Solution of the Schr\"{o}dinger
  equation by a spectral method}. \emph{J. Comp. Phys.} \textbf{1982},
  \emph{47}, 412 -- 433\relax
\mciteBstWouldAddEndPuncttrue
\mciteSetBstMidEndSepPunct{\mcitedefaultmidpunct}
{\mcitedefaultendpunct}{\mcitedefaultseppunct}\relax
\EndOfBibitem
\bibitem[Dai and Zhang(1995)Dai, and Zhang]{dai95:1491}
Dai,~J.; Zhang,~J. Z.~H. Noise-free spectrum for time-dependent calculation of
  eigenenergies. \emph{J. Chem. Phys.} \textbf{1995}, \emph{103},
  1491--1497\relax
\mciteBstWouldAddEndPuncttrue
\mciteSetBstMidEndSepPunct{\mcitedefaultmidpunct}
{\mcitedefaultendpunct}{\mcitedefaultseppunct}\relax
\EndOfBibitem
\bibitem[Mahapatra and Sathyamurthy(1995)Mahapatra, and
  Sathyamurthy]{mah95:6057}
Mahapatra,~S.; Sathyamurthy,~N. {Correlation function approach to transition
  state resonances in collinear (He,H$_2^+$) collisions}. \emph{J. Chem. Phys.}
  \textbf{1995}, \emph{102}, 6057--6066\relax
\mciteBstWouldAddEndPuncttrue
\mciteSetBstMidEndSepPunct{\mcitedefaultmidpunct}
{\mcitedefaultendpunct}{\mcitedefaultseppunct}\relax
\EndOfBibitem
\bibitem[Ibo()]{IboView}
Iboview: Orbital visualization program, G. Knizia
  \texttt{https://www.iboview.org}, Accessed: 09.04.2020\relax
\mciteBstWouldAddEndPuncttrue
\mciteSetBstMidEndSepPunct{\mcitedefaultmidpunct}
{\mcitedefaultendpunct}{\mcitedefaultseppunct}\relax
\EndOfBibitem
\bibitem[Trickl \latin{et~al.}(1989)Trickl, Cromwell, Lee, and
  Kung]{trickl.n2:1989}
Trickl,~T.; Cromwell,~E.~F.; Lee,~Y.~T.; Kung,~A.~H. State-selective ionization
  of nitrogen in the X$^2 \Sigma_g^+ (v_+ = 0)$ and $(v_+ = 1)$ states by
  two-color (1+1) photon excitation near threshold. \emph{J. Chem. Phys.}
  \textbf{1989}, \emph{91}, 6006--6012\relax
\mciteBstWouldAddEndPuncttrue
\mciteSetBstMidEndSepPunct{\mcitedefaultmidpunct}
{\mcitedefaultendpunct}{\mcitedefaultseppunct}\relax
\EndOfBibitem
\bibitem[Lide(2007)]{crchandbook:2007}
Lide,~D.~R. \emph{CRC Handbook of Chemistry and Physics}, 88th ed.; CRC Press,
  2007\relax
\mciteBstWouldAddEndPuncttrue
\mciteSetBstMidEndSepPunct{\mcitedefaultmidpunct}
{\mcitedefaultendpunct}{\mcitedefaultseppunct}\relax
\EndOfBibitem
\bibitem[Salehi \latin{et~al.}(2019)Salehi, Koner, and Meuwly]{salehi.n3m:2019}
Salehi,~S.~M.; Koner,~D.; Meuwly,~M. Vibrational Spectroscopy of N$_{3}^{-}$ in
  the gas and condensed phase. \emph{J. Phys. Chem. B} \textbf{2019},
  \emph{123}, 3282--3290\relax
\mciteBstWouldAddEndPuncttrue
\mciteSetBstMidEndSepPunct{\mcitedefaultmidpunct}
{\mcitedefaultendpunct}{\mcitedefaultseppunct}\relax
\EndOfBibitem
\bibitem[Brooks \latin{et~al.}(2009)Brooks, Brooks~III, Mackerell~Jr., Nilsson,
  Petrella, Roux, Won, Archontis, Bartels, Boresch, Caflisch, Caves, Cui,
  Dinner, Feig, Fischer, Gao, Hodoscek, Im, Kuczera, Lazaridis, Ma,
  Ovchinnikov, Paci, Pastor, Post, Pu, Schaefer, Tidor, Venable, Woodcock, Wu,
  Yang, York, and Karplus]{charmm:2009}
Brooks,~B.~R. \latin{et~al.}  CHARMM: The biomolecular simulation program.
  \emph{J. Comput. Chem.} \textbf{2009}, \emph{30}, 1545--1614\relax
\mciteBstWouldAddEndPuncttrue
\mciteSetBstMidEndSepPunct{\mcitedefaultmidpunct}
{\mcitedefaultendpunct}{\mcitedefaultseppunct}\relax
\EndOfBibitem
\bibitem[Koner \latin{et~al.}({2018})Koner, Bemish, and Meuwly]{MM.cno:2018}
Koner,~D.; Bemish,~R.~J.; Meuwly,~M. {The C($^3$P) + NO(X$^2\Pi$) $\rightarrow$
  O($^3$P) + CN(X$^2\Sigma^+$), N($^2$D)/N($^4$S) + CO(X$^1\Sigma^+$) reaction:
  Rates, branching ratios, and final states from 15 K to 20 000 K}. \emph{J.
  Chem. Phys.} \textbf{{2018}}, \emph{{149}}, {094305}\relax
\mciteBstWouldAddEndPuncttrue
\mciteSetBstMidEndSepPunct{\mcitedefaultmidpunct}
{\mcitedefaultendpunct}{\mcitedefaultseppunct}\relax
\EndOfBibitem
\bibitem[San Vicente~Veliz \latin{et~al.}({2020})San Vicente~Veliz, Koner,
  Schwilk, Bemish, and Meuwly]{MM.no2:2020}
San Vicente~Veliz,~J.~C.; Koner,~D.; Schwilk,~M.; Bemish,~R.~J.; Meuwly,~M.
  {The N($^4$S) + O$_2$(X$^3 \Sigma_g^-$) $\leftrightarrow$ O($^3$P) + NO(X$^2
  \Pi$) reaction: thermal and vibrational relaxation rates for the $^2$A$'$,
  $^4$A$'$ and $^2$A$''$ states}. \emph{Phys Chem Chem Phys.} \textbf{{2020}},
  \emph{{22}}, {3927--3939}\relax
\mciteBstWouldAddEndPuncttrue
\mciteSetBstMidEndSepPunct{\mcitedefaultmidpunct}
{\mcitedefaultendpunct}{\mcitedefaultseppunct}\relax
\EndOfBibitem
\end{mcitethebibliography}


\providecommand{\latin}[1]{#1}
\makeatletter
\providecommand{\doi}
  {\begingroup\let\do\@makeother\dospecials
  \catcode`\{=1 \catcode`\}=2 \doi@aux}
\providecommand{\doi@aux}[1]{\endgroup\texttt{#1}}
\makeatother
\providecommand*\mcitethebibliography{\thebibliography}
\csname @ifundefined\endcsname{endmcitethebibliography}
  {\let\endmcitethebibliography\endthebibliography}{}
\begin{mcitethebibliography}{0}
\providecommand*\natexlab[1]{#1}
\providecommand*\mciteSetBstSublistMode[1]{}
\providecommand*\mciteSetBstMaxWidthForm[2]{}
\providecommand*\mciteBstWouldAddEndPuncttrue
  {\def\EndOfBibitem{\unskip.}}
\providecommand*\mciteBstWouldAddEndPunctfalse
  {\let\EndOfBibitem\relax}
\providecommand*\mciteSetBstMidEndSepPunct[3]{}
\providecommand*\mciteSetBstSublistLabelBeginEnd[3]{}
\providecommand*\EndOfBibitem{}
\mciteSetBstSublistMode{f}
\mciteSetBstMaxWidthForm{subitem}{(\alph{mcitesubitemcount})}
\mciteSetBstSublistLabelBeginEnd
  {\mcitemaxwidthsubitemform\space}
  {\relax}
  {\relax}

\end{mcitethebibliography}
\end{document}


\date{\today}

\begin{table}[ht]
\caption{Grid details for the triatomic N$_3^+$ potentials. Units are
  a.u.}
\begin{tabular}{lcc}
\hline
\hline
    & \ \ \ CCSD(T)-F12b \ \ \ & \ \ \ MRCI+Q \ \ \ \\
    \hline
    $N_{R}$ & 30 & 29 \\
    $R_{\rm min}$ & 1.7 & 2.2 \\
    $R_{\rm max}$ & 10.0 & 10.0 \\
    $N_{r}$ & 18 & 22  \\
    $r_{\rm min}$ & 1.5 & 1.55 \\
    $r_{\rm max}$ & 3.1 & 4.0 \\
    $N_{\theta}$ & 13 & 7 \\
\hline
\hline
\end{tabular}
\label{sitab:triat-grid}
\end{table}

\begin{figure}[htbp]
\includegraphics[width=0.85\textwidth]{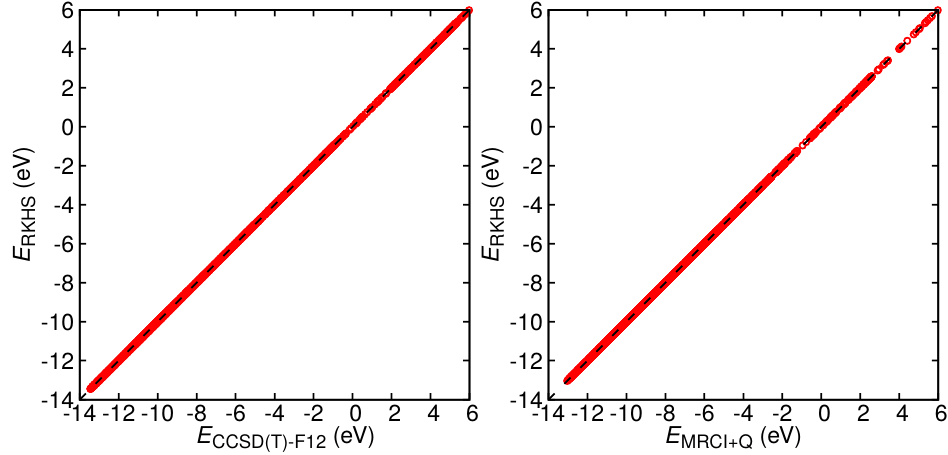}
\caption{Correlation between {\it ab initio} and RKHS energies for the
  on-grid points. The zero of energy is the $E_{\rm N}+E_{\rm
    N}+E_{\rm N^+}$ asymptote. Left panel for the CCSD(T)-F12b PES and
  right panel for the MRCI+Q PES. The respective correlation
  coefficients are $R_{\rm CCSD(T)}^{2}=1 - 4\times10^{-7}$ and
  $R_{\rm MRCI}^{2}= 1 - 3\times10^{-7}$.}
\label{sifig:grid}	
\end{figure}

\begin{figure}[htbp]
\includegraphics[width=0.85\textwidth]{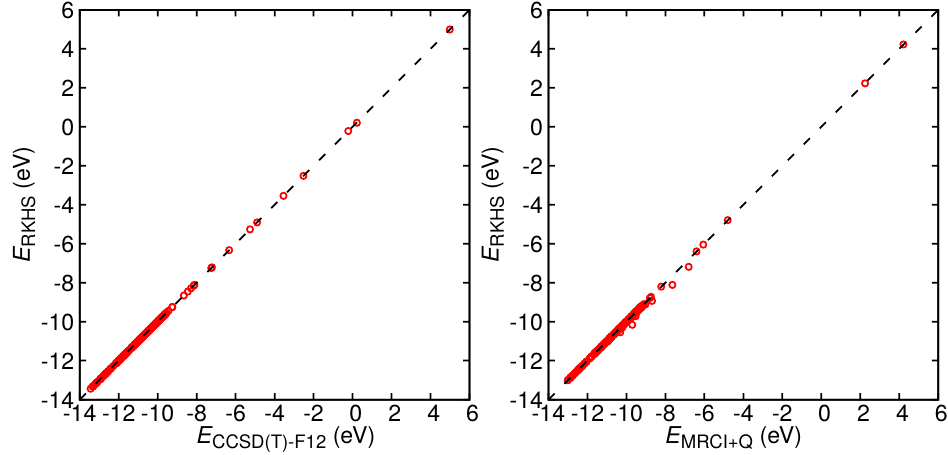}
\caption{Correlation between {\it ab initio} and RKHS-interpolated
  energies for off-grid points at the CCSD(T)-F12b (left) and MRCI+Q
  (right) levels of theory. The zero of energy is the $E_{\rm
    N}+E_{\rm N}+E_{\rm N^+}$ asymptote. The respective correlation
  coefficients are $R_{\rm CCSD(T)}^{2}=0.9999$ and
  $R_{\rm MRCI}^{2}= 0.9993$.}
\label{sifig:offgrid}	
\end{figure}

\begin{figure}[htbp]
\begin{center}
\includegraphics[width=0.8\textwidth]{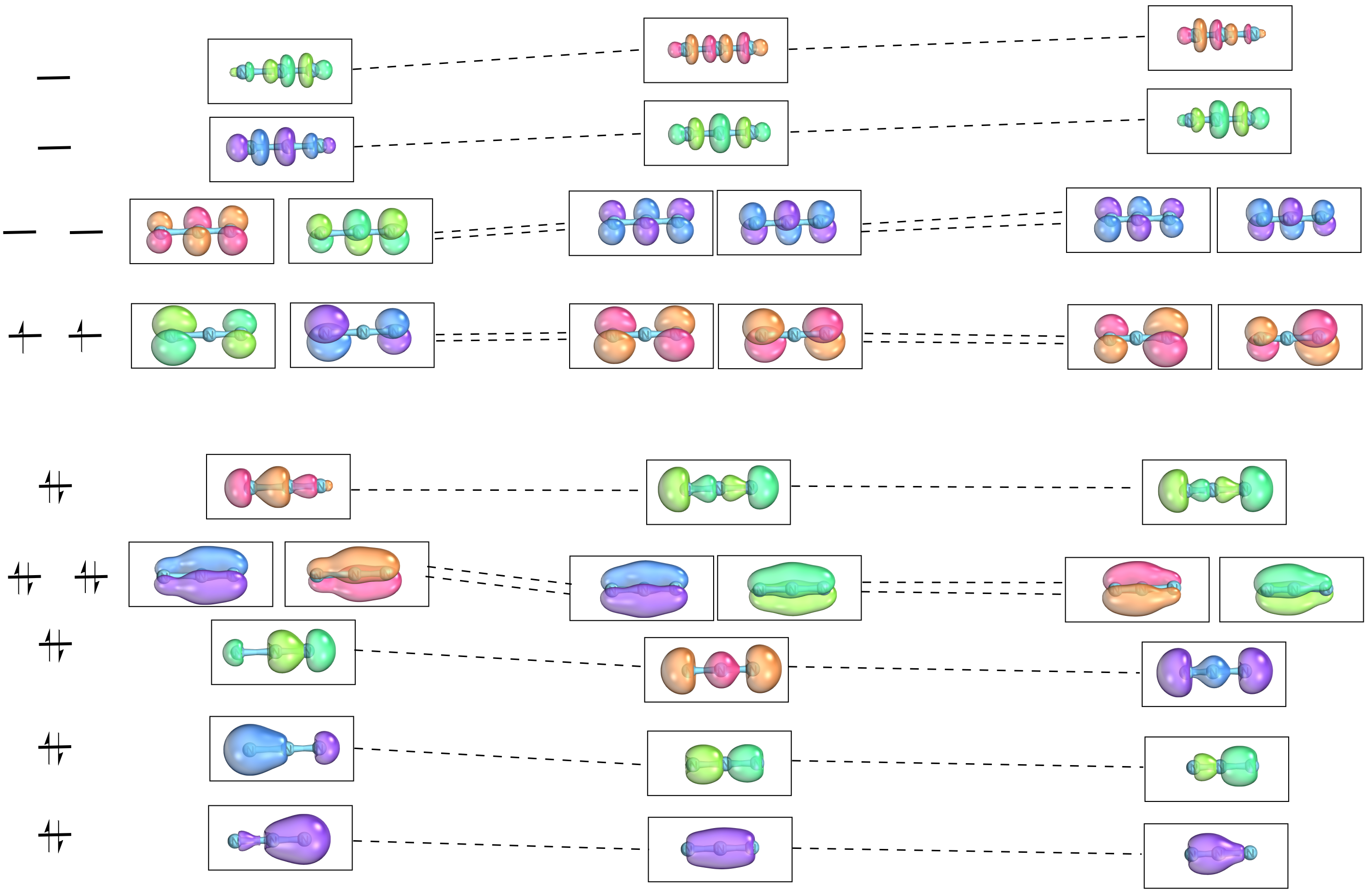}
\caption{MO diagram of the evolution of the valence space natural
  orbitals for the triplet state linear bonding of N$^+$ to N$_2$:
  Formation of covalent bonds (left panel), equilibrium structure
  (middle), and very small value of $R$.}
\label{sifig:modiag_lin_detail}
\end{center}
\end{figure}

\begin{figure}[htbp]
\begin{center}
\includegraphics[width=0.4\textwidth]{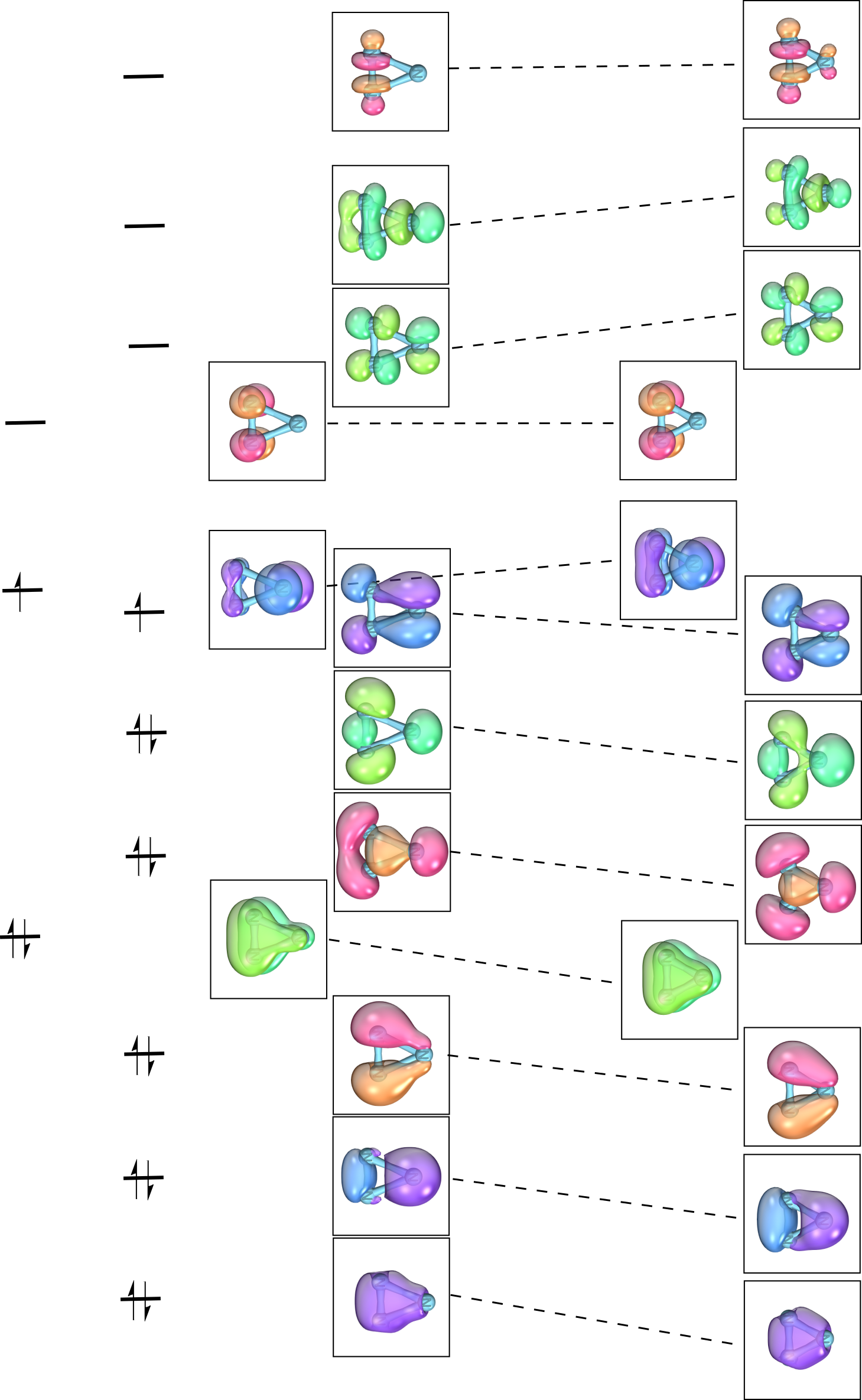}
\caption{MO diagram of the evolution of the valence space natural
  orbitals for the T-shaped bonding of N$^+$ to N$_2$: Formation of
  covalent bonds (left panel) and triangular structure with three
  nearly equidistant bonds.}
\label{sifig:modiag_tshp_detail}
\end{center}
\end{figure}

\begin{figure}[htbp]
\begin{center}
\includegraphics[width=0.99\textwidth]{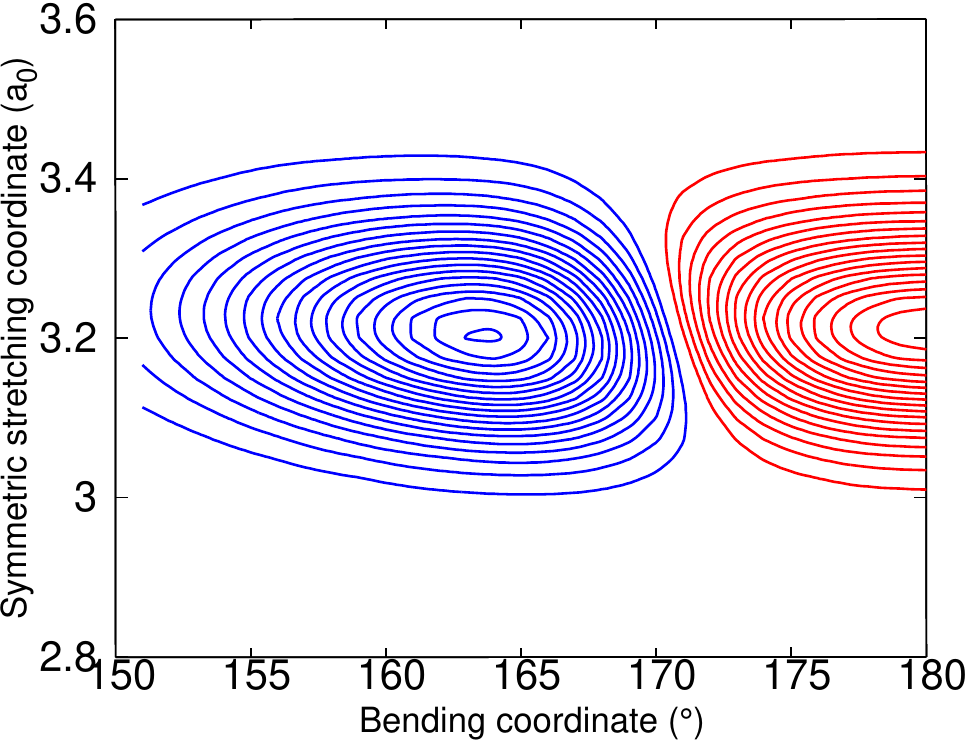}
\caption{Wavefunction for the (0 2 0 0) state.}
\label{sifig:wave0200}
\end{center}
\end{figure}

\begin{figure}[htbp]
\begin{center}
\includegraphics[width=0.99\textwidth]{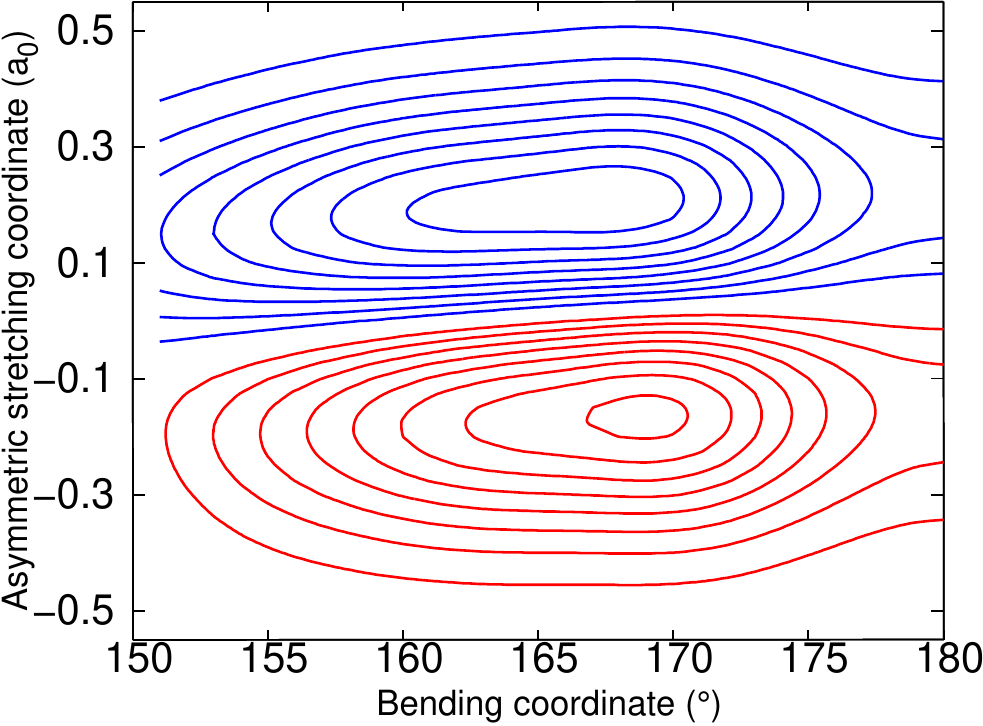}
\caption{Wavefunction for the (0 1 1 1) state.}
\label{sifig:wave0111}
\end{center}
\end{figure}

\begin{figure}[htbp]
\begin{center}
\includegraphics[width=0.99\textwidth]{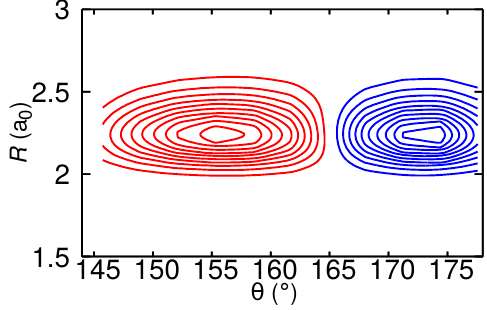}
\caption{Wavefunction for the (0 3 0 1) state.}
\label{sifig:wave0301}
\end{center}
\end{figure}

\begin{figure}[htbp]
\begin{center}
\includegraphics[width=0.99\textwidth]{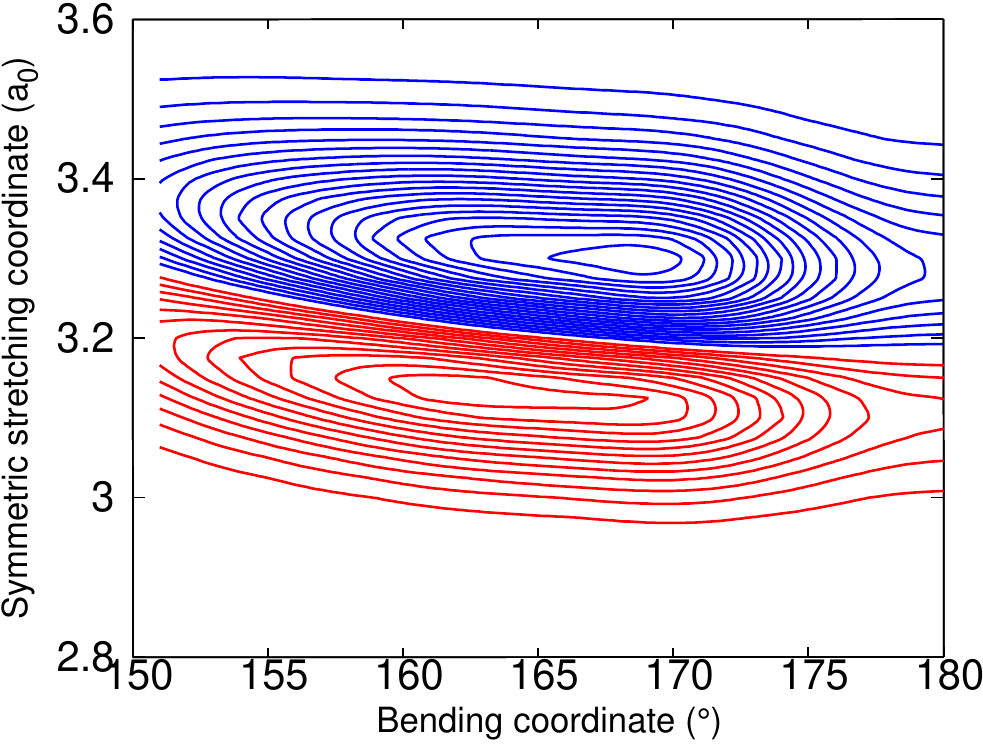}
\caption{Wavefunction for the (1 1 0 1) state.}
\label{sifig:wave1101}
\end{center}
\end{figure}

\begin{figure}[htbp]
\begin{center}
\includegraphics[width=0.99\textwidth]{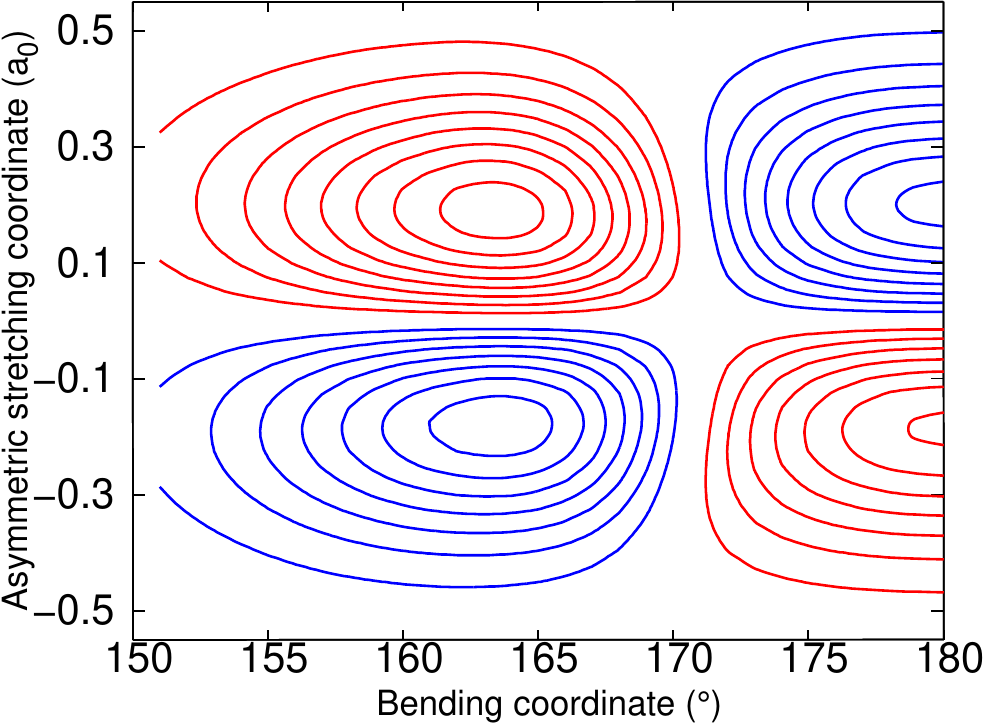}
\caption{Wavefunction for the (0 2 1 0) state.}
\label{sifig:wave0210}
\end{center}
\end{figure}

\begin{figure}[htbp]
\begin{center}
\includegraphics[width=0.99\textwidth]{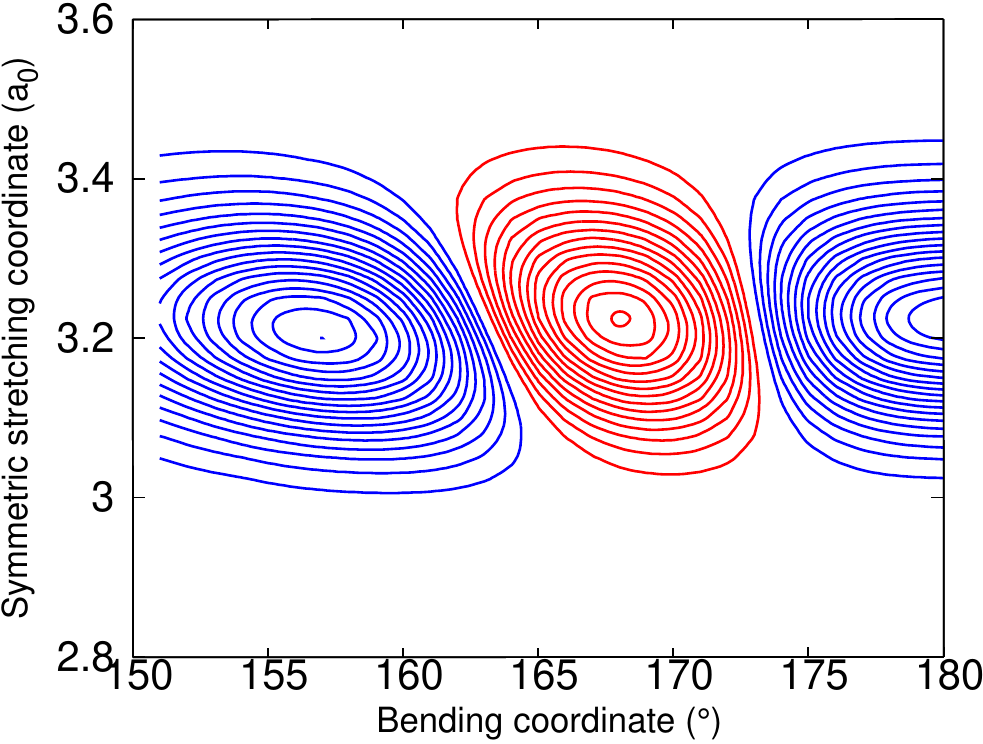}
\caption{Wavefunction for the (0 4 0 0) state.}
\label{sifig:wave0400}
\end{center}
\end{figure}
